\documentclass[5p,times]{elsarticle}

\usepackage{amssymb}
\usepackage{graphicx}

\usepackage{subcaption}
\usepackage{float}
\usepackage{amssymb}
\usepackage{threeparttable}
\usepackage{algorithm}
\usepackage{amsmath}
\usepackage{algorithmic}
\usepackage{comment}
\usepackage{booktabs}
\usepackage{multirow}
\usepackage{listings}
\usepackage{bibentry}
\usepackage{newfloat}
\usepackage[normalem]{ulem}
\useunder{\uline}{\ul}{}
\newtheorem{prof}{Proof}
\newtheorem{proposition}{Proposition}

\newcommand{\lev}{Lorentz equivariant}
\newcommand{\lec}{Lorentz equivariance}
\newcommand{\li}{Lorentz invariant}
\newcommand{\uig}{user-item graph}

\begin{document}
\begin{frontmatter}
\title{Lorentz Equivariant Model for Knowledge-Enhanced Hyperbolic Collaborative Filtering}

\author[mymainaddress]{Bosong Huang}
\ead{bosonghuang@m.scnu.edu.cn}

\author[mysecondaryaddress]{Weihao Yu}
\ead{yuwh3@chinatelecom.cn}

\author[mymainaddress]{Ruzhong Xie}
\ead{rzxie@m.scnu.edu.cn}

\author[mymainaddress]{Jing Xiao}
\ead{xiaojing@scnu.edu.cn}

\author[mymainaddress]{Jin Huang\corref{mycorrespondingauthor}}
\cortext[mycorrespondingauthor]{Corresponding author}
\ead{huangjin@m.scnu.edu.cn}

\address[mymainaddress]{South China Normal University, Guangzhou, China}
\address[mysecondaryaddress]{Research Institute of China Telecom Corporate Ltd., Guangzhou, China}

\begin{abstract}
Introducing prior auxiliary information from the knowledge graph  (KG) to assist the user-item graph can improve the comprehensive performance of the recommender system. Many recent studies show that the ensemble properties of hyperbolic spaces fit the scale-free and hierarchical characteristics exhibited in the above two types of graphs  well. 
However, existing hyperbolic methods ignore the consideration of equivariance, thus they cannot generalize symmetric features under given transformations, which seriously limits the capability of the model.
Moreover, they cannot balance preserving the heterogeneity and mining the high-order entity information to users across two graphs.
To fill these gaps, we propose a  rigorously Lorentz group equivariant knowledge-enhanced collaborative filtering model (LECF). Innovatively, we jointly update the attribute embeddings (containing the high-order entity signals from the KG) and hyperbolic embeddings (the distance between hyperbolic embeddings reveals the recommendation tendency) by the LECF layer with Lorentz Equivariant Transformation. Moreover, we propose  Hyperbolic Sparse Attention Mechanism to sample the most informative neighbor nodes. Lorentz equivariance is strictly maintained throughout the entire model, and enforcing  equivariance is proven necessary experimentally. Extensive experiments on three real-world benchmarks demonstrate that LECF remarkably outperforms state-of-the-art methods.
\end{abstract}


\end{frontmatter}

\section{Introduction}

In the era of exponential growth in information volume, collaborative filtering (CF), one of the leading technical branches of recommender systems, aims to mine similar item preferences for similar users. In knowledge-enhanced  CF, knowledge graphs (KGs) are used as auxiliary information for items to alleviate the data sparsity and cold-start problems. Recently, based on the success of graph neural networks, representations with high-order neighbor information can be learned in Euclidean space to improve recommendation performance \cite{KGCN:bm3,CKAN:bm6,KGAT:bm4,KGNN-LS:bm5}.

\begin{figure}[t]
	\centering
		\begin{subfigure}{0.45\linewidth}
			\centering
			\includegraphics[width=\linewidth]{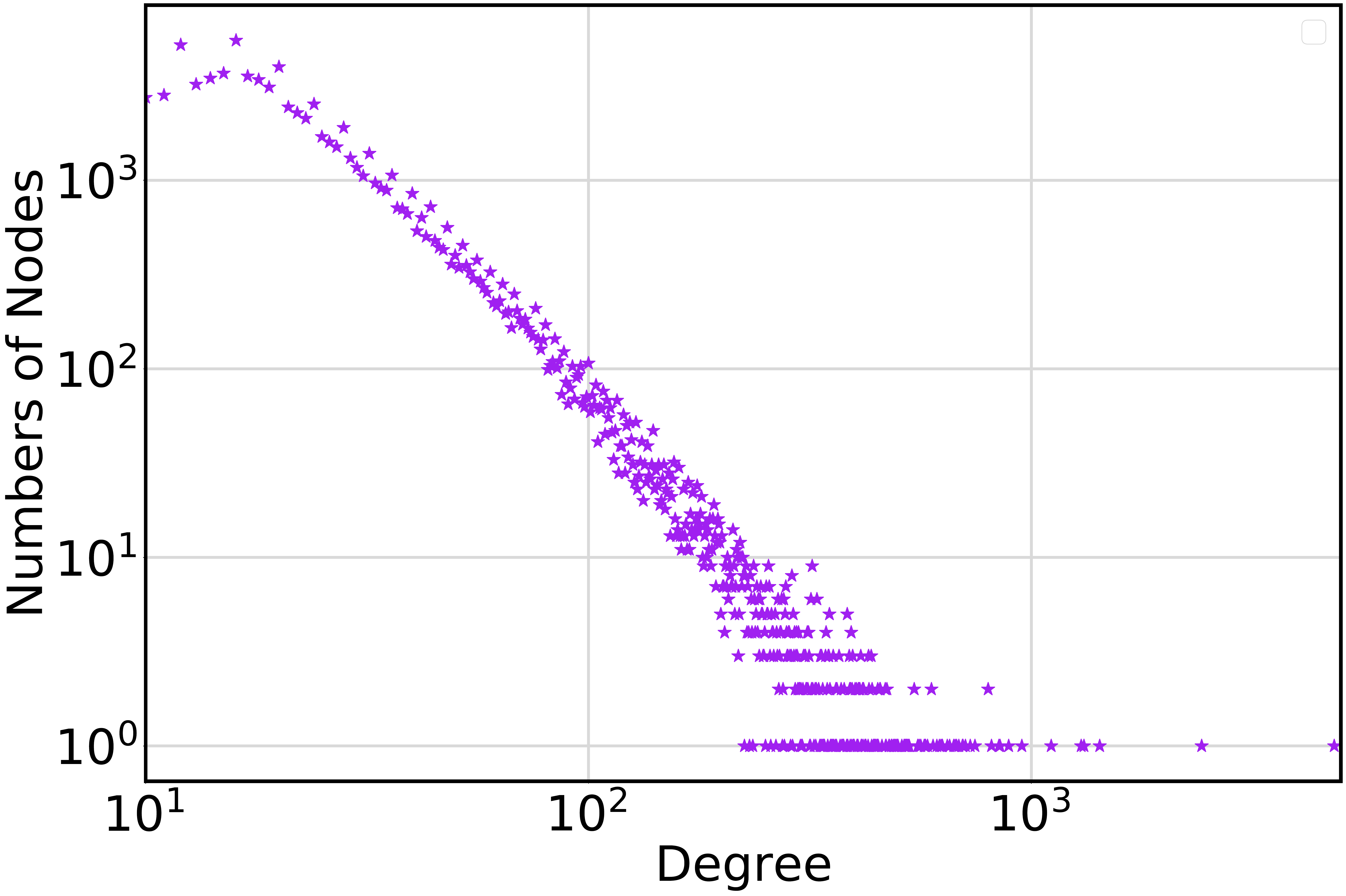}
			\caption{ Degree distribution in the \uig\ of  Book-Crossing dataset.}
			\label{fig:powerlow1}
		\end{subfigure}
		\begin{subfigure}{0.45\linewidth}
			\centering
			\includegraphics[width=\linewidth]{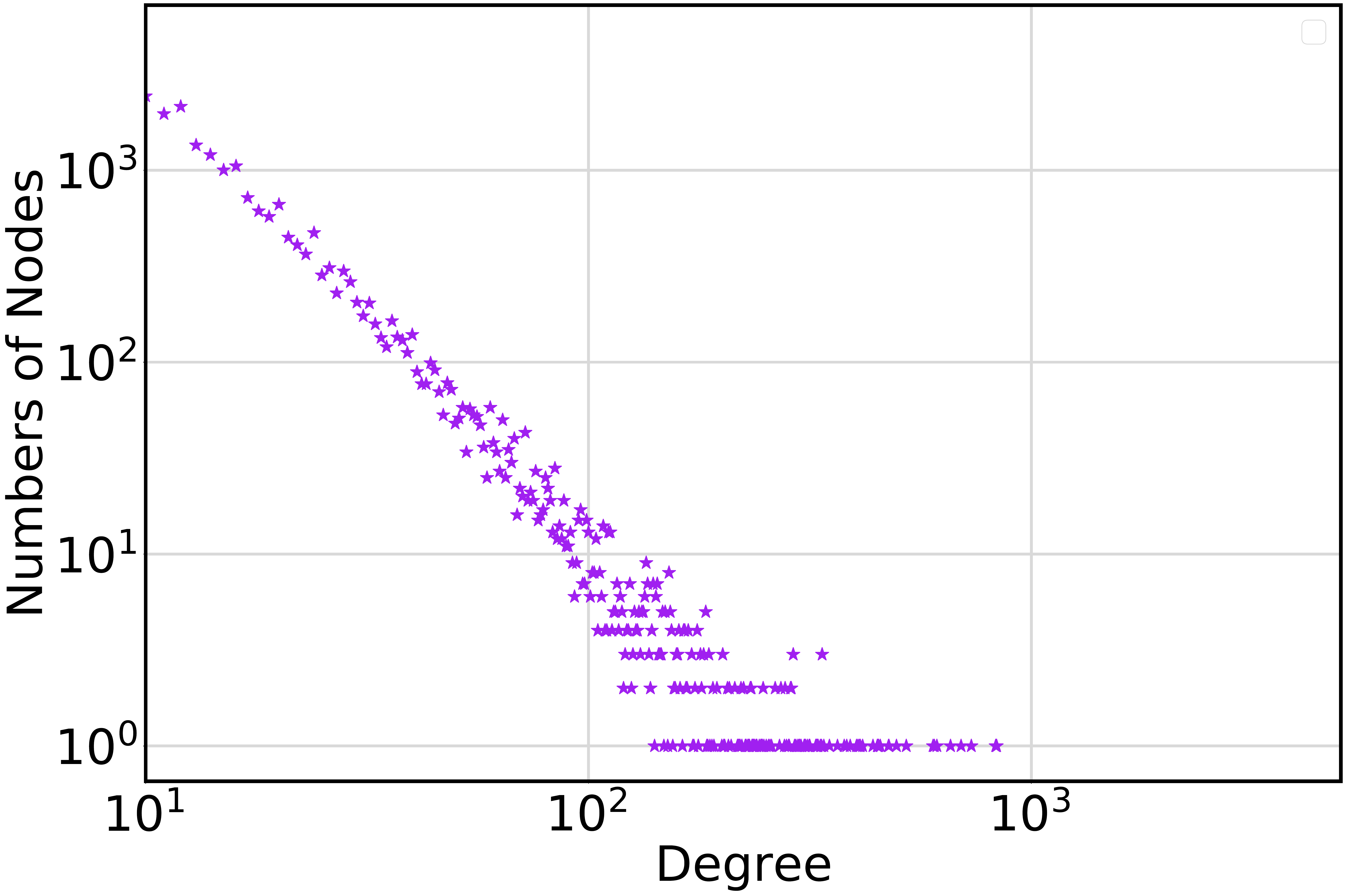}
			\caption{ Degree distribution in the knowledge graph of  Book-Crossing dataset.}
			\label{fig:powerlow2}
		\end{subfigure}
  \caption{}\label{fig:powerlow}
\end{figure}

 However, the above  Euclidean methods fail to model the significant characteristics of large-scale \uig s and KGs, e.g., \textit{scale-free (power-law distribution)}, as shown in Figure \ref{fig:powerlow} and \textit{hierarchical structures}  \cite{LKGR2022,Hgcf2021}. Unlike the homogeneous Euclidean space, the capacity of a hyperbolic space (a Riemannian manifold of negative curvature) grows exponentially as nodes move away from the origin. More importantly,  a hyperbolic can be viewed as a tree-like structure in high-dimensional space, thus fitting the hierarchy. Benefitting from the above advantages, CF models \cite{LKGR2022,HAKG2022,Hgcf2021,HICF2022,HRCF2022} based on hyperbolic representation learning achieve competitive performance in recommender systems. 

Despite the breakthroughs in hyperbolic CF, some fundamental issues have yet to be discussed and resolved, owing to the shallow understanding of hyperbolic geometry.

\begin{figure}[t]
	\centering
		\begin{subfigure}{\linewidth}
			\centering
			\includegraphics[width=0.8\linewidth]{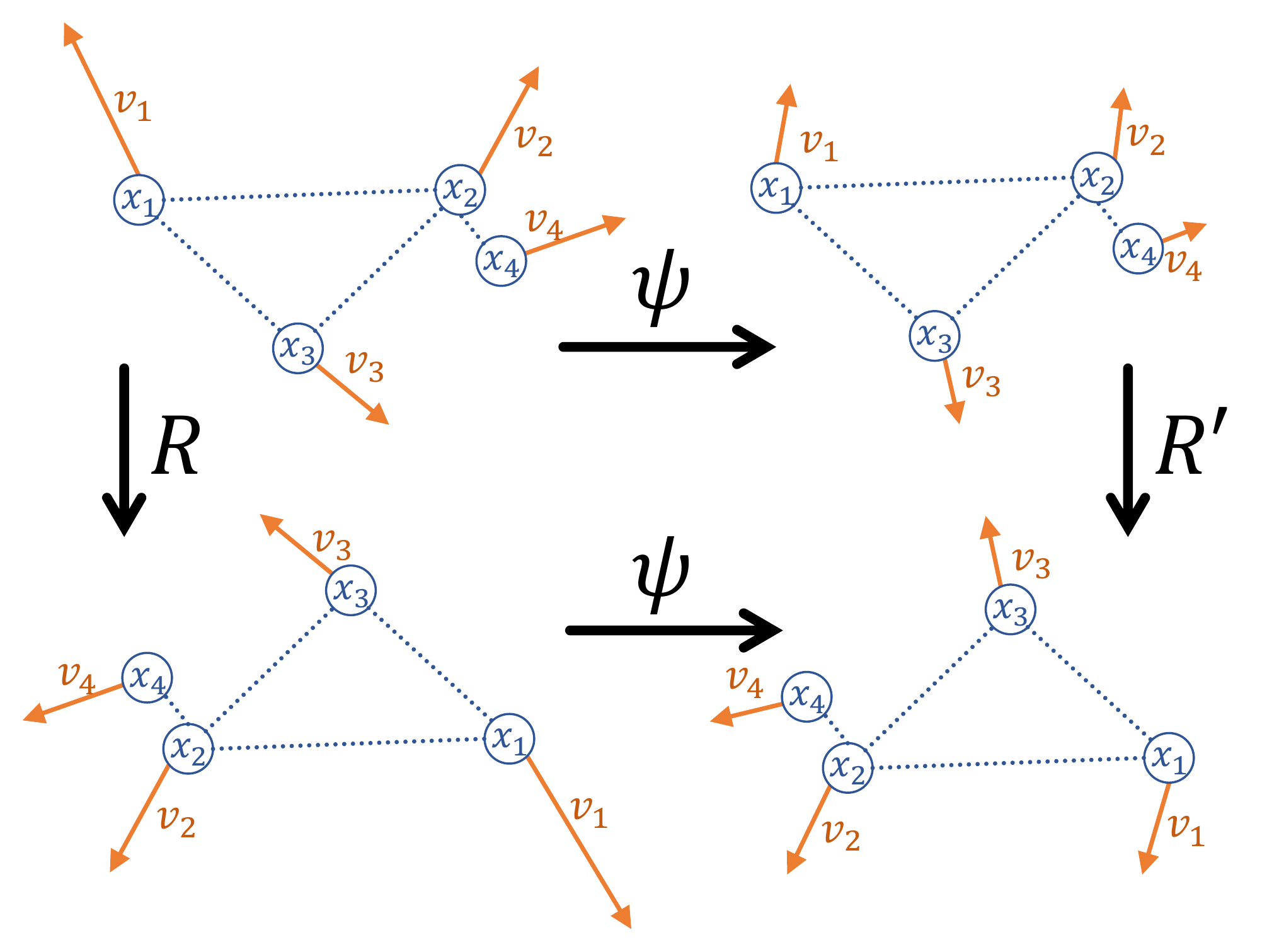}
			\caption{ Illustration of rotation equivariance on a graph respect to function $\psi$.}
			\label{fig:eqv1}
		\end{subfigure}
		\begin{subfigure}{\linewidth}
			\centering
			\includegraphics[width=\linewidth]{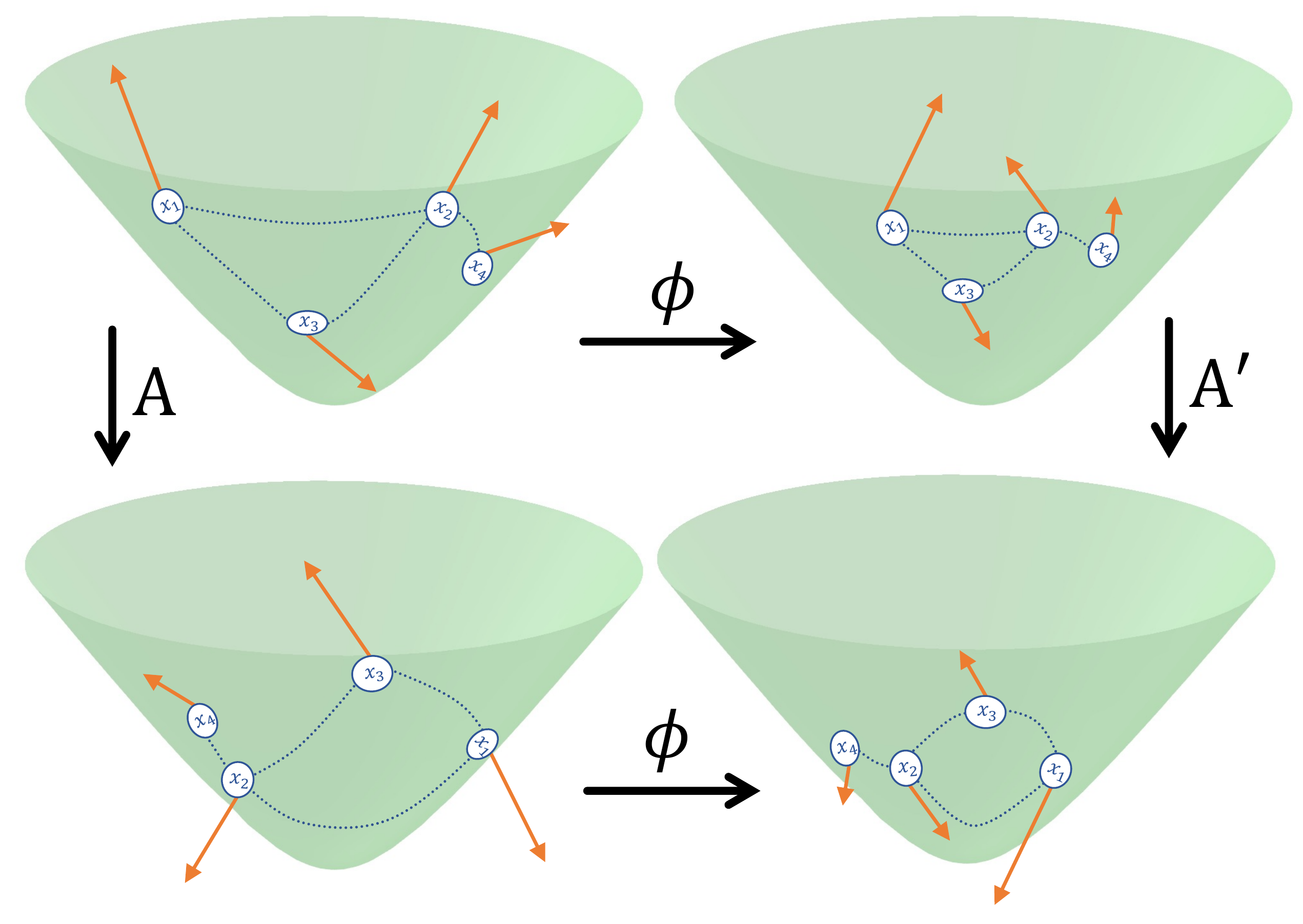}
			\caption{ Illustration of Lorentz equivariance in the hyperbolic space respect to function $\phi$.}
			\label{fig:eqv2}
		\end{subfigure}
  \caption{}
\end{figure}

\cite{PEG2021equivariant,ER2022er,deepER} state the importance of enforcing Euclidean equivariance in the social network or KG representation learning; that is, the features the model captured should  be full-fidelity
 under any transformations since the relative relation in those representations should not be distorted under any transformation (for example, in Figure \ref{fig:eqv1}, the Euclidean equivariant function $\psi$ generalizes the same graph representation under different rotation transformations).
 The existing hyperbolic algorithm aims to perform recommendation ranking through the hyperbolic distance of the hyperbolic representation. At the same time, the symmetric representations in the hyperbolic space have more severe distortions than those in Euclidean space \cite{FullyHNN2021,NestedHNN2022}. Therefore, the fidelity of the hyperbolic CF model with respect to symmetry features (i.e., maintaining  Lorentz equivariance) is much more urgent to guarantee than in Euclidean space (as shown in Figure \ref{fig:eqv2}, the Lorentz equivariant function $\phi$ perceives the same features under different Lorentz transformations).
 However, existing  hyperbolic CF algorithms ignore this. Specifically, they apply message aggregation by projecting the representations into the tangent space. Unfortunately, this aggregation method breaks the equivariance in the hyperbolic space (i.e., Lorentz equivariance), and the distortion accumulates as the aggregation layers are staked. In addition, they arbitrarily perform linear transformations on hyperbolic representations, which is also one of the sources of breaking equivariance. The potential of hyperbolic geometry for recommender systems is yet to be fully exploited due to the disregard of \lec.
 
\begin{figure}[t]
			\centering
\includegraphics[width=\linewidth]{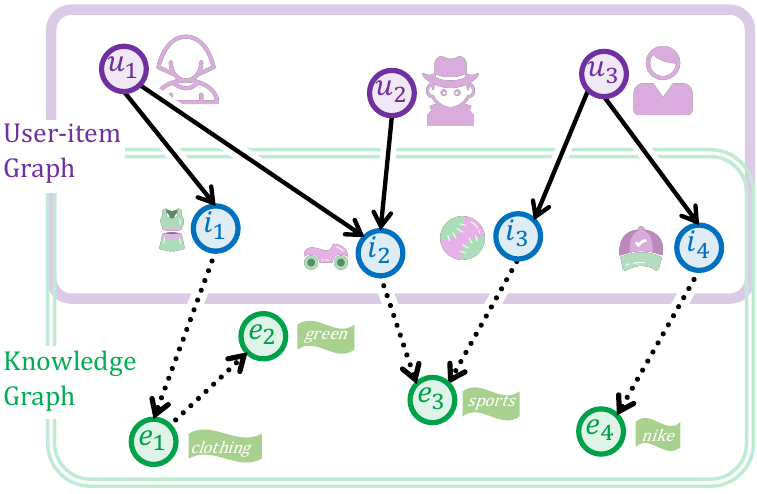}
			\caption{ User-item graph with auxiliary knowledge graph.}
			\label{fig:graphpath}
\end{figure}

 Although an auxiliary KG enhances hyperbolic CF,   preserving the heterogeneity of KG and \uig\ while performing high-order CF between them is challenging.  A straightforward way \cite{LKGR2022} is to perform different message passing processes for the two graphs separately, which completely ignores the high-order entity signals across two graphs to users (e.g., in Fig \ref{fig:graphpath}, the path across two graphs to users: $ e_4 \rightarrow i_4  \rightarrow u_3 \rightarrow e_3  \rightarrow e_3  \rightarrow i_2  \rightarrow  u_1  $ ). Another idea \cite{HAKG2022} is to use gated aggregation for two graphs in the hyperbolic space, which is too complicated and computationally expensive. Therefore, a concise and effective framework that responds to the above issues is urgently needed.

In response, we propose  \textbf{L}orentz \textbf{E}quivariant \textbf{C}ollaborative \textbf{F}iltering (LECF) model for knowledge-enhanced recommendation, which is divided into two stages. \textbf{Attribute generation stage}: First, Hyperbolic Sparse Attention Mechanism is proposed to sample the most informative neighbor nodes. Next, when obtaining the higher-order entity representation in KG by the proposed Item Attribute Generator (IAG), we employ the hyperbolic distance centroid strategy rather than the previous tangent space aggregation strategy, which breaks \lec.
 \textbf{CF stage}: The high-order entity representations obtained from the first stage are input to LECF layers as the item attributes.   To pass the high-order entity signals to users across graphs, the node attribute and hyperbolic embeddings are mutually updated by the well-designed structure of LECF layer and proposed  Lorentz Equivariant Transformation (carrying the node attribute signals). Each component of our model strictly maintains Lorentz equivariance, and the potential of hyperbolic geometry is thus stimulated. Our contributions are summarized below:
\begin{itemize}
\item[$\bullet$] LECF inaugurates the study of equivariance in hyperbolic CF. Specifically, LECF  can better perceive the symmetric characteristics under the Lorentz transformations by enforcing \lec, thereby stimulating the potential generalization capability.

\item[$\bullet$] Through the collaboration of IAGs and  LECF layers with Lorentz Equivariant Transformations, we skillfully balance preserving the heterogeneity and mining the high-order entity information to users across graphs.  

\item[$\bullet$] We propose Hyperbolic Sparse Attention Mechanism to replace the random sampling of the KG and \uig\ for selecting the most informative neighbor nodes. 

\item[$\bullet$]  We conduct extensive experiments on three public benchmark datasets to demonstrate the groundbreaking superiority of LECF.
    
\end{itemize}

\section{Related Work}

\subsection{Knowledge-Enhanced Collaborative Filtering}

The introduction of KGs enhances the recommendation in a two-fold way. That is, it provides rich semantic information for items and additional information dissemination paths out of the \uig. The most straightforward way is to match the KG and the \uig\ into a unified graph for representation learning \cite{CKAN:bm6,KGAT:bm4,MKGAT}. 
However, these methods focus more on learning representations of the unified graph and ignore the heterogeneity between the \uig\ and KG. Aiming at these problems, the meth-path based methods \cite{MCRec,TMER} manually define the high-order information propagation path in the recommendation task but seriously rely on more domain expert knowledge. Another idea is to use \uig\ as the auxiliary information for the deep relationship mining in KG while simplifying the exploration of collaborative information between users and items \cite{KGNN-LS:bm5,KGCN}. \cite{entity2rec,Atbrg} further apply graph pre-training and rule constraints to generate subgraphs that meet application requirements, thereby providing support for relationship mining between users and items.

Since \cite{Nickel2017,Nickel2018,HyperbolicneuralnetworksGanea,HyperbolicAttentionNetworks}, representation learning for fitting the latent hierarchy in data in hyperbolic space has become popular.
HGCN \cite{HyperbolicGNN2019Leskovec} further combines hyperbolic representation learning with the message passing network, projecting the node representations from the hyperbolic space to the tangent space for message aggregation. Based on the success of the above studies, \cite{LowDimensionalHyperbolicKG,hyperbolicattentionKG,KnowledgeAssociationKG,hyperboliconesKG} find that the hyperbolic space can well fit the hierarchical and logical patterns in KGs, and thus achieve a series of competitive performances through the high-fidelity and concise representation learning.
Similarly, \uig s also exhibit the characteristics of a hierarchical and power-law distribution, and \cite{Hme2020,Hyperml2020} therefore demonstrate the significance of combining hyperbolic geometry with recommendation tasks. \cite{Hgcf2021,HICF2022,HRCF2022} capture higher-order information in user-item graphs by incorporating multiple layers of hyperbolic message passing networks.

Based on the successful practice of hyperbolic geometry in recommendation tasks and KG representation learning, \cite{LKGR2022,HAKG2022} further attempt to use hyperbolic CF to address the problem of knowledge-enhanced recommendation, as
the high-fidelity and compact potential of hyperbolic spaces well fit complex relation paradigms in such tasks. Although the auxiliary KG improves recommendation performance, the potential of hyperbolic geometry is not fully exploited since many components are superficially migrated from the Euclidean space.

\subsection{Exploration of Equivariance}
\label{sec:ExplorationofEquivariance}
Different tasks require specific generalizations for various symmetric features in the data. Thus, a series of studies focuses on exploring the diverse equivariance and invariance of neural networks for corresponding tasks. 

The most straightforward case is that in image processing, \cite{Understandingimage2015,Harmonicnetworks} hope that the model can ultimately retain the data features after translating, rotating, and reflecting the image, that is, the equivariance of the model in the Euclidean space (for example, the model should be able to recognize the cat in the image with equal probability no matter under which the above transformations). With the rise of graph neural networks, \cite{fastGNN2016,Semi-supervisedGNN2016,sage2017inductive} claim that  the permutation equivariance is a basic principle. Compared with general graph data, geometric graph (i.e., graph nodes that contain spatial coordinates) data exhibit a higher demand for equivariance of the model, as the physical and chemical properties and geometric relationships behind them will not change due to spatial transformation. 
Specifically, \cite{E(n)GNN2021} propose E(n) equivariant graph neural networks, which achieve leading results in molecular property prediction. In the tasks of computational physics and chemistry, \cite{improveEGNN} generalize E(3) equivariant graph networks so that the nodes can contain covariant information, such as vectors or tensors. \cite{xugeodiff,Equivariantdiff} further introduce the discussion of geometric equivariance in molecular and protein generation models. 
The exploration of equivariance is far from limited to Euclidean space. For example, \cite{Generalizing2020lie} enforces equivariance to transformations from any specified Lie group in the vector space. Additionally, \cite{practical2021,Scalars2021} turn their attention to equivariant models under other groups, e.g., SO(n), O(1, 3), Sp(n), and SU(n).
Equivariance has also been discussed by emerging researches in representation learning for social networks or recommendations. For example, \cite{PEG2021equivariant} designs a rotation equivariant position encoding representation learning method to address the isomorphism problem of network structures. \cite{ER2022er} improves the generalization ability by adding semantic equivariance constraints in knowledge graph representation learning. \cite{EquivariantContrastiveLearning} introduces equivariant contrastive learning into data augmentation for the sequential recommendation.

As an emerging field, hyperbolic representation learning is still in the ascendant for its study of equivariance (i.e., Lorentz equivariance). \cite{practical2021,jet2022,Autoencoders2022lorentz}, from the perspective of physics, shows that the models achieve a significant advantage by enforcing Lorentz equivariance. However, another majority of studies based on the Lorentz model still urgently need to study the problem of equivariance.

\begin{figure}[t]
	\centering
		\begin{subfigure}{0.85\linewidth}
			\centering
			\includegraphics[width=\linewidth]{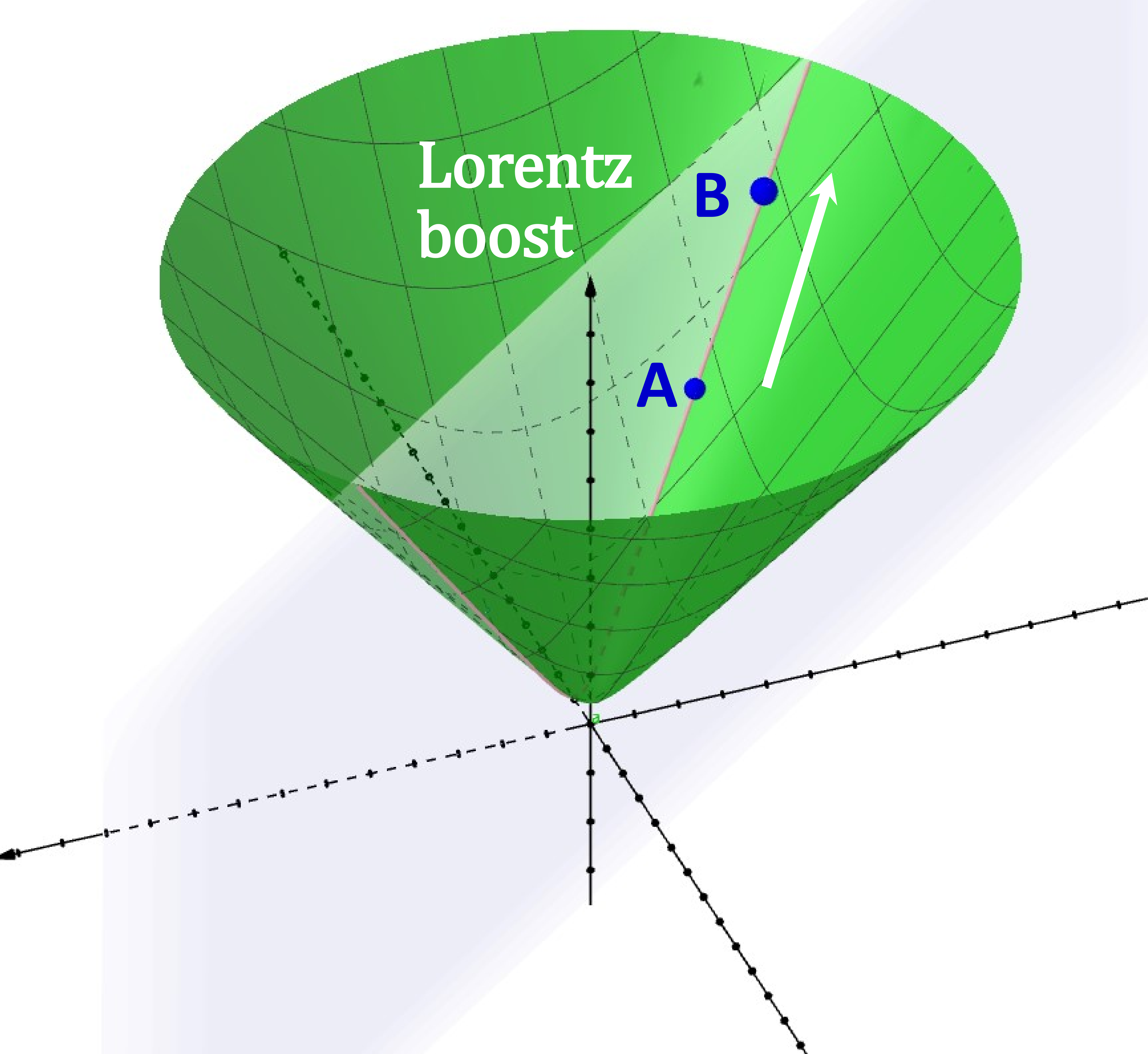}
			\caption{ Lorentz boost.}
			\label{fig:2a}
		\end{subfigure}
		\begin{subfigure}{0.85\linewidth}
			\centering
			\includegraphics[width=\linewidth]{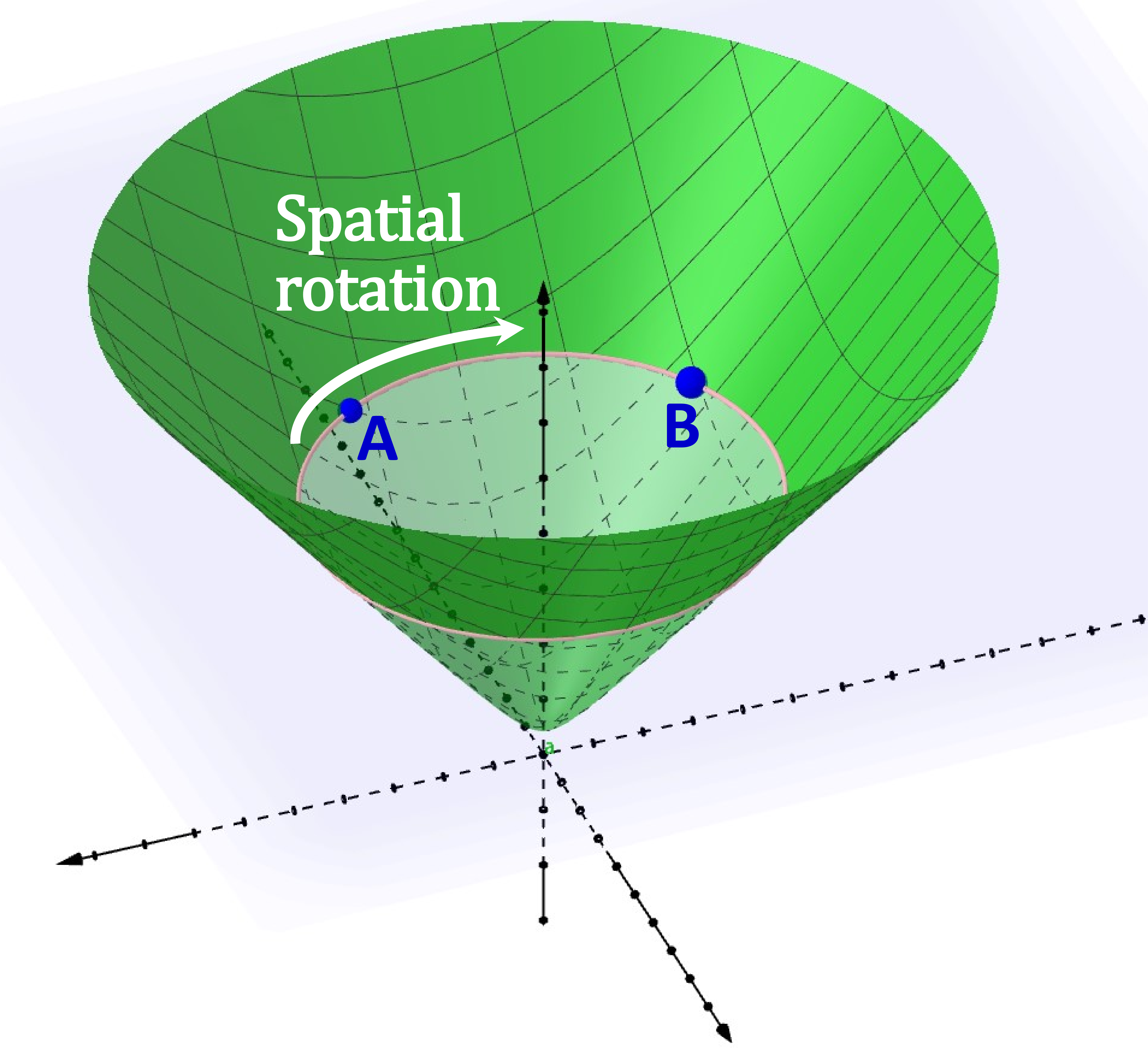}
			\caption{Spatial rotation.}
			\label{fig:2b}
		\end{subfigure}
  \caption{ Two types of Lorentz transformations.}\label{fig:2}
\end{figure}

\section{Preliminary}
\subsection{Hyperbolic Geometry}
In this paper, the n-dimensional Euclidean space is denoted by $\mathbb{R}^n$. An n-dimensional hyperbolic space is a Riemannian manifold of negative curvature $-1/C(C>0)$. For numerical stability, we choose one of the equivalent models in the hyperbolic space, the Lorentz model, which is defined  as $(\mathbb{H}^n,g_\mathcal{L})$. Where $\mathbb{H}^{n}=\left\{\mathbf{x} \in \mathbb{R}^{n+1}:\langle\mathbf{x}, \mathbf{x}\rangle_{\mathcal{L}}=-C, x_{0}>0\right\}$ and $g_{\mathcal{L}}$ is the metric tensor that is defined as $g_{\mathcal{L}}=diag[-1,1,1,\dots,1]$.  $\langle\mathbf{x}, \mathbf{x}\rangle_\mathcal{L}$ is the Lorentz inner product, which is elaborated below:
\begin{equation}
\langle\mathbf{x}, \mathbf{y}\rangle_{\mathcal{L}} =-x_{0} y_{0}+\sum_{i=1}^{n} x_{i} y_{i}
\end{equation}
where $\mathbf{x}, \mathbf{y} \in \mathbb{H}^{n}$. Based on this,  we can give the definition of the distance between two points: 
\begin{equation}
d_{\mathcal{L}}(\mathbf{x}, \mathbf{y})=\sqrt{C} \operatorname{arcosh}\left(-\langle\mathbf{x}, \mathbf{y}\rangle_{\mathcal{L}} / C\right)
\end{equation}
The Lorentz norm can thus be defined as:
\begin{equation}
    \| \mathbf{x}\|_{\mathcal{L}}=\sqrt{\langle\mathbf{x}, \mathbf{x}\rangle_{\mathcal{L}}}
\end{equation}

\subsection{Lorentz Model}
In the application of the Lorentz model to special relativity, the first dimension $\mathbf{x}_0$ is interpreted as the time axis, and the remaining dimensions $\mathbf{x}_{1 \sim n}$ are interpreted as the space axes \cite{jet2022}. For ease of description, we inherit these terminologies.
\subsubsection{Lorentz Transformation.}
In the Lorentz model, the linear isometries are called the Lorentz transformation, which could further be polar decomposed \cite{polarDecomposition2002} into a combination of a Lorentz boost and a spatial rotation; see Figure \ref{fig:2} for details. 
A polar decomposition for a  Lorentz transformation $\mathbf{A} \in \mathbf{SO}^{+}(1,n)$\footnote{$\mathbf{SO}^{+}(1,n)$ is the positive special Lorentz group} is defined as
\begin{equation}
\mathbf{A}=\left[\begin{array}{ll}
1 & 0 \\
0 & \mathbf{R}
\end{array}\right]\left[\begin{array}{ccc}
\mathrm{cosh \alpha} &\mathrm{sinh \alpha} & 0\\
\mathrm{sinh \alpha} & \mathrm{cosh \alpha} &0\\ 0&0& I_{n-1}
\end{array}\right]
\end{equation}
where $\mathbf{R} \in \mathbf{SO}(n)$\footnote{$\mathbf{SO}(n)$ is the special orthogonal matrix}, $\alpha$ is the hyperbolic angle, and $I_{n - 1}$ is the $(n - 1)\times(n - 1)$
identity matrix. The left half is the spatial rotation, and the right half is the Lorentz boost. 
\subsubsection{Lorentz Equivariance.}
We first present the general concept of equivariance that encompasses all cases in Section \ref{sec:ExplorationofEquivariance}. $T$ is an arbitrary transformation of $\mathbf{x} \rightarrow \mathbf{x}$ in the abstract group $g$. A function $\psi(\cdot)$ is equivariant to $g$ when it satisfies the following conditions:

\begin{equation}
    \psi(T(\mathbf{x}))=T'(\psi(\mathbf{x}))
\end{equation}
where $T' \in g$ is the equivalent transformation of $T$. From this we can give the related concepts of Lorentz equivariance.
Let $\mathbf{A}$ be the Lorentz transformation in the Lorentz group and $\mathbf{v} \in \mathbb{H}^n$, the Lorentz equivariance of
$\phi (\cdot)$ means there exists an equivalent Lorentz transformation $\mathbf{A'}$ such that:
\begin{equation}
\mathbf{A}\phi(\mathbf{v})=\phi(\mathbf{A'}\mathbf{v})\qquad where \quad\mathbf{v},\phi(\mathbf{v}) \in \mathbb{H}^n
    \label{equ:equva1}
\end{equation}
Lorentz invariance of $\phi (\cdot)$ is thus defined as:
\begin{equation}
    \phi(\mathbf{v})=\phi(\mathbf{A}\mathbf{v})\qquad where\quad \phi(\mathbf{v}) \in \mathbb{R}^n
    \label{equ:equva2}
\end{equation}

\begin{table}[t]
\caption{Symbols and their definitions.}\label{table:symbols}
\begin{tabular}{ll}
\midrule[1.5pt] Symbol & Definition \\
\midrule[0.3pt]

$\mathbb{R}^n$&N-dimension Euclidean space   \\ 
$\mathbb{H}^n$& N-dimension hyperbolic space  \\ 
$d_{\mathcal{L}}(,)$ &The hyperbolic distance between two nodes \\ 
$\|\cdot\|_{\mathcal{L}}$&The Lorentz norm  \\ 
$-1/C(C>0)$ & Curvature of the  hyperbolic space\\

$\mathbf{e}_i \in \mathbb{H}^n $ & Hyperbolic embedding of a KG entity \\
$\mathbf{n}_i \in \mathbb{H}^n $ & Hyperbolic embedding of node $n_i$ \\
$\mathbf{r}^k$ & Relation embedding of KG \\
${a}(i, j)$ &  The hyperbolic attention coefficient\\
$\hat{a}(i, j)$ &  The hyperbolic sparse attention coefficient\\ 

$\mathbf{x}_u \in \mathbb{H}^n$ & Hyperbolic embedding of a user \\
$\mathbf{x}_i \in \mathbb{H}^n$ & Hyperbolic embedding of an item \\
$\mathbf{h}_u \in \mathbb{R}^n$ & Attribute embedding of a user \\
$\mathbf{h}_i \in \mathbb{R}^n$ & Attribute embedding of an item \\

\midrule[1.5pt]
\end{tabular}
\end{table}

\section{Methodology}

\subsection{Problem Formulation}
In this subsection, we clarify the data structure used in the study and the task definition.
\subsubsection{User-item Graph.} The user-item graph (interaction) represents the collection of the behaviors of user $u$ on item $i$, denoted by $
\left\{\left(u, r^{u}, i\right) \mid u \in \mathcal{U}, i \in \mathcal{I}\right\}
$, where $r^u$ is the abstract representation of user behavior (e.g., clicking, purchasing, and rating).
The user-item interaction matrix $\mathbf{Y}$ is a binary matrix, abstracted by setting a certain threshold for user-item graph interactions ($\mathbf{Y}_{ui}$= 1  indicates that there is an interaction; otherwise, $\mathbf{Y}_{ui}$ = 0).
Items in the user-item graph can be matched with the entities in the KG.
\subsubsection{Knowledge Graph (KG).} A KG is a collection of triples representing real-world concepts and their relationships: $\left\{\left(e_{i}, r^k, e_{j}\right) \mid e_{i}, e_{j} \in \mathcal{E}, r^k \in \mathcal{R}\right\}$. Each triple $\left(e_{i}, r^k, e_{j}\right)$ indicates that the head entity $e_{i}$ has a relationship $r_k$ that points to the tail entity $e_j$. There are many studies proving that KGs can improve recommendation performance by providing semantic and attribute information. Here, we introduce a KG to generate item attributes for the user-item graph.
\subsubsection{Task Definition.} Given a user-item graph and a KG, the task studied in this paper is to provide the predicted recommendation score $\hat{y}_{ui}$ that the user $u$ adopts  item $i$ in the test set.

\subsection{Overview of the Model}
In summary, our model can be summarized into two stages. In the attribute generation stage, we abandon the random neighbor node sampling strategy while proposing  Hyperbolic Sparse Attention Mechanism to sample the most informative neighbor nodes. Then, the proposed Item Attribute Generator (IAG), takes the hyperbolic distance centroids as the aggregated neighbor information to generate item attributes.  In the CF stage, the entity representations obtained from IAG are input into LECF layers as the item attributes of \uig. In each layer, the node attribute and  hyperbolic embeddings are mutually updated by manipulating  Lorentz Equivariant Transformations. Last but not least, we theoretically prove that our model strictly preserves the Lorentz equivariance.

\subsection{Hyperbolic Sparse Attention Mechanism}
Existing CF algorithms mostly perform random sampling when selecting neighbor nodes, thus ignoring the importance levels of neighbors. Instead, we combine the sampling process and attention mechanism,  proposing Hyperbolic Sparse Attention Mechanism to sample the most informative neighbor nodes. Specifically, the more uniformly the mechanism distributes the attention coefficients (i.e., the smaller the information entropy of the attention coefficients), the less effective it is \cite{Informer2021}. Therefore, we adopt information entropy to measure the difference between the attention coefficient and the uniform distribution to select more informative neighbor nodes. In short, we select  $t$ nodes corresponding to the query with the smallest information entropy of the attention coefficients in the hyperbolic space. Inspired by \cite{Hattention2018}, we calculate the attention coefficient $a(i,j)$ between the  node $\mathbf{n}_i \in \mathbb{H}^n$ and its neighbors $\mathbf{n}_j \in \mathcal{N}(\mathbf{n}_i) $: 
\begin{equation}
\begin{aligned}
\overline{a}(i, j)&=\mathrm{exp}\left(\mathbf{W}_{ij} d_{\mathcal{L}}\left(\mathbf{n}_i, \mathbf{n}_j\right)\right) \\
a(i,j)&=\frac{{\overline{a}(i, j)} \gamma(\mathbf{n}_i)}{\sum_{\mathbf{n}_{j'} \in \mathcal{N}(\mathbf{n}_i)} { \overline{a}(i, j')}\gamma(\mathbf{n}_{j'})} 
\end{aligned}\label{equ:hba1}
\end{equation}
where $\mathbf{W}$ is the parameter matrix and  $\gamma(\mathbf{n})=\frac{1}{\sqrt{1-\|\mathbf{n}\|^2}}$  is the Lorentz factor. 
Then we calculate the probability distribution of the attention coefficients on each neighbor node. Finally,  the importance measure $m(j)$ is calculated by: 
\begin{equation}
\begin{aligned}
\mathrm{P}(i\mid j)&=\frac{\mathrm{exp}({{a}(i, j)})}{\sum_{\mathbf{n}_{i'} \in \mathcal{N}(\mathbf{n}_j)} \mathrm{exp}({ {a}(i', j)})}
\\m(j)&=-\sum_{\mathbf{n}_{i'} \in \mathcal{N}(\mathbf{n}_j)} \mathrm{P}\left(i'\mid j\right) \log \mathrm{P}\left(i'\mid j\right)
\end{aligned}\label{equ:hba2}
\end{equation}
For each node $\mathbf{n}_i$, we select its top $t$ neighbor nodes $\mathbf{n}_j \in \mathcal{N}(\mathbf{n}_i) $ with the smallest $m(j)$ and fill the blanks with zeros to derive the final sparse attention coefficient $\hat{a}(i, j)$.

It is worth noting that, Equation \ref{equ:hba1} is the generalized definition of the attention coefficient. For the hyperbolic embedding $\mathbf{e}$  that represents the KG entities and the relation embeddings $\mathbf{r}^k$, we replace Equation 7 with:
\begin{equation}
\begin{aligned}
\overline{a}(i,r^k,j)&=\mathrm{exp}\left(d_{\mathcal{L}}\left(f_{\mathbf{r}^k}(\mathbf{e}_i), \mathbf{e}_j\right)\right) \\
a(i,r^k,j)&=\frac{{\overline{a}(i,r^k, j)} \gamma(\mathbf{e}_i)}{\sum_{\mathbf{e}_{j'} \in \mathcal{N}(\mathbf{e}_i)} { \overline{a}(i, r^k,j')}\gamma(\mathbf{e}_{j'})} 
\end{aligned}\label{equ:hbaKG}
\end{equation}

where $f_{\mathbf{r}^k}(\cdot)$ is Lorentz Equivariant Transformation  parameterized by the relation $\mathbf{r}^k$, see Section \ref{sec:LorentzEquivariantTransformation} for the specific definition.

\subsection{Item Attribute Generator (IAG)}
The item attributes are generated from the higher-order representations of the entities in the KG that match them. In IAG, the entity representations are initialized in the hyperbolic space by the method \cite{Hgcf2021}, denoted as  $\mathbf{e}^{0}$. We perform message propagation on entity representations in the hyperbolic space. To avoid the feature distortion and \lec\ destruction caused by aggregation in tangent space, we adopt the algorithm \cite{LGCN2021WWW} that uses the hyperbolic distance centroid as the aggregation result. The optimization objective is:
\begin{equation}
    \arg \min _{\mathbf{e}_{i}^{l+1} \in \mathbb{H}^{n}} \sum_ {\mathbf{e}^{l}_j\in \mathcal{N}(\mathbf{e}_i)} \hat{a}(i, r^k,j) d_{\mathcal{L}}\left( \mathbf{e}_{i}^{l+1},\mathbf{e}_j^{l}\right)
    \label{equ:iag1}
\end{equation}
Equation \ref{equ:iag1} has a closed-form solution:
\begin{equation}
\mathbf{e}_{i}^{l+1}=\sqrt{C} \frac{\sum _{\mathbf{e}_j\in \mathcal{N}(\mathbf{e}_i)} \hat{a}(i, r^k,j) \mathbf{e}_{j}^{l}} {|\|\sum _{\mathbf{e}_j\in \mathcal{N}(\mathbf{e}_i)}   \hat{a}(i, r^k,j) \mathbf{e}_{j}^{l} \|_{\mathcal{L}} |}    \label{equ:iag2}
\end{equation}
After stacking several IAG layers, the obtained entity representations will carry the high-order information in the KG.
\subsection{Lorentz Equivariant Collaborative Filtering  (LECF) Layer}
Before implementing the CF, 
we match the aggregated entity representations from IAG  with the items in \uig, taking them as the initial item attribute embeddings, denoted by $\mathbf{h}^0_i$.  Moreover, the user attribute embeddings $\mathbf{h}^0_u$ are initialized as ones. Then, we follow the method in \cite{Hgcf2021} to initialize the user and item hyperbolic embeddings, denoted by $\mathbf{x}^{0}_u$ and $ \mathbf{x}^{0}_i$.

To preserve the heterogeneity between the \uig\ and KG while mining the high-order entity information to users across graphs,  we formulate LECF layer to update the node attribute and  hyperbolic embeddings jointly as follows\footnote{We take the updating of user representations as an instance, for the items are the same.}:
\begin{equation}
    {{\mathbf{m}}_{ui}}={{\phi }_{e}}\left(\mathbf{h}_{u}^{l},\mathbf{h}_{i}^{l},{{ d_{\mathcal{L}}\left( \mathbf{x}_{u}^{l},\mathbf{x}_{i}^{l}  \right)}}\right)
\label{equ:lecf1}
\end{equation}

\begin{equation}
\mathbf{x}_{u}^{l+1}=\sqrt{C} \frac{\sum _{i\in \mathcal{N}(u)} \hat{a}\left(u, i\right) {\pi}\left({f_{\mathbf{x}_i}(\mathbf{m}_{ui})},\mathbf{x}_{i}^{l}\right)} {|\|\sum _{i\in \mathcal{N}(u)}        \hat{a}\left(u,i\right) \pi\left({f_{\mathbf{x}_i}(\mathbf{m}_{ui})},\mathbf{x}_{i}^{l}\right) \|_{\mathcal{L}} |}   \label{equ:lecf2}
\end{equation}

\begin{equation}
    {\mathbf{m}_{u}}=\sum _{i\in \mathcal{N}(u)}{\mathbf{m}_{ui}}
    \label{equ:lecf3}
\end{equation}

\begin{equation}
    \mathbf{h}_{u}^{l+1}={{\phi }_\mathbf{h}}\left(\mathbf{h}_{u}^{l},{\mathbf{m}_{u}}\right)
    \label{equ:lecf4}
\end{equation}
where $\phi_{e}$ and $\phi_{\mathbf{h}}$ are the edge and node operations, respectively, implemented by multilayer perceptrons (MLPs).
 $\pi(f_{\mathbf{x}}(\mathbf{m}),\cdot)$ is the Lorentz Equivariant Transformation that  will be elaborated on in the next section.
In particular, Equation \ref{equ:lecf1} and Equation \ref{equ:lecf2} are related to hyperbolic geometry. Concretely, Equation \ref{equ:lecf1} obtains the message embeddings $\mathbf{m}_{ui}$ from the hyperbolic distances and the corresponding attributes, thus involving the information from the KG. 
In Equation \ref{equ:lecf2}, we update the user hyperbolic embeddings by the neighbor item hyperbolic embeddings manipulated by the Lorentz Equivariant Transformation (involving the message embedding $\mathbf{m}_{ui}$) to fuse the item attribute information.
In turn,  the updating  of attribute embeddings is also affected by the hyperbolic embeddings via message embeddings, according to Equation \ref{equ:lecf4}.
In order to maintain the geometric properties, both the input and output hyperbolic embeddings in Equation \ref{equ:lecf2}  are strictly embedded in the hyperbolic space; see the next section for proof. 
\subsubsection{Lorentz Equivariant Transformation.}
\label{sec:LorentzEquivariantTransformation}
Existing hyperbolic recommender systems all project the hyperbolic node embeddings to tangent space and then perform the linear transformation on Euclidean space. However, this method breaks \lec\ and does not fully release the potential of hyperbolic space. That is, it only includes spatial rotation in the Lorentz transformation and excludes the Lorentz boost \cite{FullyHNN2021}. Inspired by \cite{NestedHNN2022}, our Lorentz transformation matrix is formulated as:
\begin{equation}
f_{\mathbf{x}}(\mathbf{m})=\left[\begin{array}{c}
\frac{\sqrt{\left\|\mathbf{W}_{\mathbf{m}} \mathbf{x}\right\|^{2}+C}}{\mathbf{v}^{\top} \mathbf{x}} \mathbf{v}^{\top}\\
\mathbf{W}_{\mathbf{m}}
\end{array}\right]
\end{equation}
where $\mathbf{W}_{\mathbf{m}} \in \mathbb{R}^{n \times(n+1)}$ is transformed $\mathbf{m}$ from MLP, $\mathbf{v} \in \mathbb{R}^{1 \times(n+1)} $ is the learnable parameter vector and $f_{\mathbf{x}}(\mathbf{m})\in \mathbb{R}^{(n+1) \times(n+1)}$.
To enforce Lorentz equivariance, we propose Lorentz Equivariant Transformation to manipulate $\mathbf{x}$ by $f_{\mathbf{x}}(\mathbf{m})$ as follows:
\begin{equation}
\begin{aligned}
    \pi(f_{\mathbf{x}}(\mathbf{m}),\mathbf{x})&=\frac{f_{\mathbf{x}}(\mathbf{m})\mathbf{x}}{\|f_{\mathbf{x}}(\mathbf{m})\mathbf{x}\|_\mathcal{L}}\qquad 
    \\s.t. \, f_{\mathbf{x}}(\mathbf{m})g_\mathcal{L}\left(f_{\mathbf{x}}(\mathbf{m})\right)^{\top}&=g_\mathcal{L}
    \label{equ:pi}
\end{aligned}
\end{equation}

\begin{proposition}
$\forall \mathbf{x} \in \mathbb{H}^n \cap \mathbf{W}_{\mathbf{m}} \in \mathbb{R}^{n \times(n+1)} \cap \mathbf{v} \in \mathbb{R}^{1 \times(n+1)}$, we have $ \forall f_{\mathbf{x}}(\mathbf{m})\mathbf{x} \in \mathbb{H}^n$.
\label{Proposition:1}
\end{proposition}

\begin{prof}
$\forall \mathbf{x} \in \mathbb{H}^n $, we get $f_{\mathbf{x}}(\mathbf{m})\mathbf{x}=\tiny{\left[\begin{array}{c}
{\sqrt{\left\|\mathbf{W}_{\mathbf{m}} \mathbf{x}\right\|^{2}+C}}\\
\mathbf{W}_{\mathbf{m}}\mathbf{x}
\end{array}\right]}$. Then, we have $\langle{f_{\mathbf{x}}(\mathbf{m})\mathbf{x}},\\ {f_{\mathbf{x}}(\mathbf{m})\mathbf{x}}\rangle_\mathcal{L}=-C$, therefore $f_{\mathbf{x}}(\mathbf{m})\mathbf{x} \in \mathbb{H}^n$.
\end{prof}

According to Proposition \ref{Proposition:1}, we prove that Equation \ref{equ:lecf2} guarantees that the updating and transformation of node hyperbolic embeddings are strictly embedded in the hyperbolic space. More importantly, since $\mathbf{W_m}$ is transformed from the message embedding $\mathbf{m}_{ui}$, the entity information in the KG will guide the Lorentz transformation process of $f_\mathbf{x}$. Meanwhile, a certain degree of freedom on the time axis is retained by introducing the learnable vector $\mathbf{v}$.

\subsection{Model Prediction and Optimization}
\subsubsection{Model Prediction.}
Before the CF process,  $L_1$ layers of aggregation   (Equation  \ref{equ:iag2}) in  IAG will be stacked to propagate the high-order entity information in the KG, generating the item attributes. Then, we stack the $L_2$ layers of  LECF layers to obtain the complementary high-order hyperbolic embeddings of users  and items.  The representations of each layer in the above process will be summed up to compose the final complementary information:
\begin{equation}
    \mathbf{e}_i=\sqrt{C}\frac{\sum^{L_1}_{l=1}\omega_1^l\mathbf{e}_i^l}{|\|\sum^{L_1}_{l=1}\omega_1^l\mathbf{e}_i^l\|_{\mathcal{L}}|}\label{equ:predction1}
\end{equation}
\begin{equation}
   \mathbf{x}_u=\sqrt{C}\frac{\sum^{L_2}_{l=1}\omega_2^l\mathbf{x}_u^l}{|\|\sum^{L_2}_{l=1}\omega_2^l\mathbf{x}_u^l\|_{\mathcal{L}}|} \, ,\quad \mathbf{x}_i=\sqrt{C}\frac{\sum^{L_2}_{l=1}\omega_2^l\mathbf{x}_i^l}{|\|\sum^{L_2}_{l=1}\omega_2^l\mathbf{x}_i^l\|_{\mathcal{L}}|}\label{equ:predction2}
\end{equation}
where $\omega$ are the hyperparameters. Equation \ref{equ:predction1} corresponds to IAG, and Equation \ref{equ:predction2} corresponds to LECF. Note that the results of the summation are strictly embedded in the hyperbolic space, which was unsatisfied in previous methods \cite{Hgcf2021,HAKG2022}. Finally,  we calculate the exponential of the negative hyperbolic distance between the user and item hyperbolic embeddings as the predicting recommendation score:  $\hat{y}_{ui}=\exp \left(-d_{\mathcal{L}}\left(\mathbf{x}_{u},\mathbf{x}_{i}\right)\right)$.

\subsubsection{Model Optimization.}
For a user $u$, let $i^+$ be the positive item, i.e., $\mathbf{Y}_{u,i^+}=1$. Let $i^-$ be the negative item, i.e., $\mathbf{Y}_{u,i^-}=0$. In this paper,  we sample one positive and one negative item for a user. The optimization goal  is to make the user approximate the positive item  and move away from negative items in the hyperbolic space; thus, the loss function is defined as:
\begin{equation}
Loss=\sum_{u \in \mathcal{U}}\sum_{i \in \mathcal{N}(u)} \mathrm{max}\left(  \hat{y}_{u, i^-}-\hat{y}_{u, i^+}+m,0\right)+\lambda\|\Theta\|_{2}^{2}
\label{equ:loss}
\end{equation}
where $m$ is the separation margin and $\|\Theta\|_{2}^{2}$ is the $\mathrm{L2}$ regularizer assigned by coefficient $\lambda$.

\subsection{Analysis of Lorentz Equivariance}
In this section, we analyze the Lorentz equivariance of the entire model, i.e., the two stages strictly satisfy one of Equations \ref{equ:equva1}, \ref{equ:equva2} and Proposition \ref{prop:2}.
\begin{proposition}
\cite{Scalars2021} A continuous function $\phi(\cdot)$ is Lorentz equivariant  if and only if 
$\phi\left(\mathbf{v}_{1}, \mathbf{v}_{2}, \cdots, \mathbf{v}_{N}\right)=\sum_{i=1}^{N} g_{i}\left(\left\langle \mathbf{v}_{i}, \mathbf{v}_{j}\right\rangle_{\mathcal{L}}^{N}\right) \mathbf{v}_{i}$, where $g_{i}(\cdot)$ can be taken to be polynomial.
\label{prop:2}
\end{proposition}
First, we consider all the scalars and vectors in Euclidean space to be \li. For Hyperbolic Sparse Attention Mechanism, the only part that involves hyperbolic vectors is Equation \ref{equ:hba1}, and $d_{\mathcal {L}}(\cdot)$ is \li, so the entire attention mechanism is \li. For the message aggregation, i.e., Equation \ref{equ:iag2} in IAG, we can easily prove that it is \lev\ according to Proposition \ref{prop:2}, and because $\hat{a}(\cdot)$ is \li, IAG is thus \lev.
 For LECF layer, since $d_{\mathcal {L}}(\cdot)$ is \li, Equation \ref{equ:lecf1} is \li. In Equation \ref{equ:lecf2}, $\pi\left({f_{\mathbf{x}_i}(\mathbf{m}_{ui})},\mathbf{x}_{i}^{l}\right)$ is \lev\ according to \cite{NestedHNN2022}. Moreover, the aggregation process  is \lev\ as discussed above; thus, the entire LECF layer is \lev .  Inductively,  a composition of IAGs and LECF layers will also be \lev . Finally, the summation of the results of each layer is \lev\ according to Proposition \ref{prop:2}, and the computation of the predicting recommendation score is \li.

\section{Experiments}
\subsection{Datasets}

We employ three benchmark datasets with auxiliary KGs to evaluate the effectiveness of LECF: Book-Crossing\footnote{ http://www2.informatik.uni-freiburg.de/~cziegler/BX/}, MovieLens-20M\footnote{https://grouplens.org/datasets/movielens/20m/}, and Yelp2018\footnote{https://www.yelp.com/dataset}. They are widely used in recent research on similar tasks and vary in application domain, data volume, and sparsity. In detail,  the auxiliary KG from each \uig\ is constructed by considering the triplets that involve two-hop neighbor entities of items. To simulate the implicit feedback setting, we convert the interactions into binary preferences by applying a threshold $\geq$ 4. To ensure KG quality, we follow the preprocess in previous work \cite{KGAT:bm4,LKGR2022}, filtering out infrequent entities (i.e., fewer than 10 in both graphs). We split the datasets into training, validation, and testing sets at a ratio of  6:2:2. The composition statistics of each dataset are shown in Table \ref{table:st1}.
\begin{table}[t]
\caption{Statistics of the datasets in the experiments.}\label{table:st1}
\centering\setlength\tabcolsep{2pt}\renewcommand{\arraystretch}{1}
\begin{tabular}{cc|c|c|cl}
\midrule[1.5pt]
\multicolumn{1}{l}{}                                                                            & \multicolumn{1}{c|}{} & \multicolumn{1}{c|}{Book-Crossing} & \multicolumn{1}{c|}{MovieLens-20M} & \multicolumn{1}{c}{Yelp2018}  \\ \midrule[0.3pt]
\multicolumn{1}{c}{\multirow{3}{*}{\begin{tabular}[c]{@{}c@{}}\end{tabular}}} & Users                 & 17,860                    & 138,159                    & 45,919                       &  \\
\multicolumn{1}{c}{}                                                                           & Items                 & 14,967                    & 16,954                     & 45,538                       &  \\
\multicolumn{1}{c}{}                                                                           & Interactions          & 139,746                   & 13,501,622                 & 1,185,068                   \\ \midrule[0.3pt]
\multicolumn{1}{c}{\multirow{3}{*}{\begin{tabular}[c]{@{}c@{}}\end{tabular}}} & Entities              & 77,903                    & 102,569                    & 90,961                       &  \\
\multicolumn{1}{c}{}                                                                           & Relations             & 25                        & 32                         & 42                           &  \\
\multicolumn{1}{c}{}                                                                           & KG triples            & 151,500                   & 499,474                    & 1,853,704                     \\ \midrule[1.5pt]
\end{tabular}
\end{table}

\begin{table*}[ht]
\caption{Recall and NDCG  results for all datasets.}\label{table:rs1}
\centering
\begin{tabular}{c|cccc|cccc|cccc}
\midrule[1.5pt]
           & \multicolumn{4}{c|}{Book-Crossing} & \multicolumn{4}{c|}{MovieLens-20M}  & \multicolumn{4}{c}{Yelp2018}  \\ 
&\tiny{R@10(\%)}&\tiny{N@10(\%)}  &\tiny{R@20(\%)}&\tiny{N@20(\%)}     &\tiny{R@10(\%)}&\tiny{N@10(\%)}  &\tiny{R@20(\%)}&\tiny{N@20(\%)}       &\tiny{R@10(\%)}&\tiny{N@10(\%)}  &\tiny{R@20(\%)}&\tiny{N@20(\%)}     \\ \midrule[0.3pt]
BRP     &2.91&2.23    & 4.50             & 2.76       &13.04&9.71      & 20.53          & 15.84     &3.82&3.51   & 6.24          & 4.29          \\
CKE     &3.01&1.85    & 4.32             & 2.31       &13.93&10.22       & 21.54          & 15.77   &4.11&3.80     & 6.52          & 4.31          \\
KGCN    &5.86&5.28    & 7.65             & 5.90       &13.36&10.39    & 19.28          & 13.45      &4.29&3.77     & 6.31          & 4.39          \\
KGAT    &4.32&2.62    & 5.29             & 3.05       &{\ul14.75}   &{\ul11.37}     & 22.26         & 17.23     &4.62&4.10    &   7.02   &  4.57    \\
KGNN-LS &7.19&5.47     & 8.50             & 5.95      &13.71&10.23    & 20.06          & 15.30      &4.10&3.74    & 6.79          & 4.20          \\
CKAN    &3.34&2.91    & 7.37             & 5.85       &13.64&10.38    & 21.33          & 15.21      &4.04&3.59    & 6.41          & 4.38         \\
Hyper-Know &4.91&2.85   & 7.17             & 5.64     &12.20&9.15    & 22.72          & 16.83       &4.21&4.02     & 6.85          & 4.46          \\
HAKG    &7.33&{\ul5.50}    & 8.70         & 6.04      &13.18&10.53    & 23.73          & 17.54      &\textbf{4.86}&\textbf{4.35} &  {\ul7.65}    & {\ul5.16}          \\
LKGR    &{\ul7.80}& 5.07    &{\ul 9.23}   &{\ul 6.72} &11.89&10.75  & {\ul 25.17}    & {\ul 20.35}  &4.57&4.22  & 6.83          & 4.35          \\ \midrule[0.3pt]
LECF    &\textbf{8.23}&\textbf{5.68}  &\textbf{10.07}  &\textbf{7.08}         &\textbf{15.62}&\textbf{11.87}  & \textbf{28.27} &\textbf{22.74}        &{\ul4.60}&{\ul4.25} & \textbf{8.32} & \textbf{5.52} \\ \midrule[0.3pt]
\% Improv. &5.5\%&3.27\% & 9.10\%           & 5.27\%           &5.90\%&4.40\%  & 12.35\%       &11.75\%                &N/A&N/A   &  8.76\%              &  6.97\% \\ \midrule[1.5pt]          
\end{tabular}
\hspace{15pt}\begin{tablenotes}\centering
 \item[1] \qquad \quad\scriptsize{The best-performing models on each dataset and metric are highlighted in \\  \, \quad \qquad bold, and the second-best  models are underlined.}
\end{tablenotes}
\end{table*}

\subsection{Experiment Settings}
\subsubsection{Baselines.}
To verify the effectiveness of LECF, we compare it with  state-of-the-art methods, including both the well-known Euclidean baselines and leading hyperbolic models. For Euclidean baselines, the KG-free method BPR is included. Furthermore, we include the mainstream propagation-based methods: \textbf{KGCN} \cite{KGCN:bm3}, \textbf{KGNN-LS} \cite{KGNN-LS:bm5}, and \textbf{CKAN} \cite{CKAN:bm6}. Additionally, regularization-based methods are included such as \textbf{CKE} \cite{CKE:bm2} and \textbf{KGAT} \cite{KGAT:bm4}. For the hyperbolic models, state-of-the-art methods with auxiliary KGs are considered: \textbf{Hyper-Know} \cite{Hyper-Know:bm7}, \textbf{HAKG}  \cite{HAKG2022}, and \textbf{LKGR} \cite{LKGR2022}.

\subsubsection{Experiment Details.}
In evaluating the Top-$K$ recommendation task, we use the trained model to rank $K$ items for each user in the test set with the highest predicted recommendation score $\hat{y}_{ui}$. Thus, two widely-used evaluation protocols \cite{recall2018}, i.e., Recall@$K$ and NDCG@$K$ ($K$= 20 by default) are employed. For a fair comparison,  the size of ID embeddings is set to 64, and we initialize the model parameters by Xavier initializer.  
 We optimize all the Euclidean models with Adam optimizer 
and the hyperbolic models with  Riemannian SGD \cite{StochasticRiemannian}. 

We train all the models on a single NVIDIA GeForce RTX 3090 GPU. Here, a grid search is used to find all optimal parameters. The search range for the regularization coefficient $\lambda$ is $\{ 10^{-2},10^{-3},10^{-4},10^{-5},10^{-6}\}$, and the aggregation order search range is from 2 to 8. The learning rate is tuned within $\{5\times10^{-2},10^{-2},5\times10^{-3},10^{-3}\}$. The range of Riemannian SGD weight decay is  $\{10^{-2},10^{-3},10^{-4},10^{-5}\}$. The curvature $-1/C$ is set constant to $-1$ in all experiments.
\subsection{Overall Performance}
The empirical results on Recall@$K$ and NDCG@$K$  ($K$=10 or $K$=20 )     are reported in Table \ref{table:rs1}. Overall, LECF significantly outperforms all Euclidean and hyperbolic models on each dataset. Furthermore, we have the following observations:  (i) the hyperbolic models are generally better than the Euclidean models due to their better compatibility with large-scale networks. However, since they do not maintain \lec, their performance is not stable across different datasets. LECF addresses this shortcoming, thus comprehensively surpassing its counterparts.
 (ii)  LKGR and HAKG outperform Hyper-Know among the hyperbolic models because they consider the high-order knowledge in hyperbolic space. In contrast, LECF further considers high-order entity signals to users across graphs, further improving performance. (iii) LECF has a more significant improvement on MovieLens-20M than the other datasets. A possible reason is that MovieLens-20M has more symmetric features than the other datasets (due to its larger data size). Since LECF can generalize better to symmetric features that benefit from enforcing \lec, the performance is more prominent.

\subsection{Validation Experiments of Lorentz Equivariance}
\begin{figure}[t]
\begin{minipage}[t]{0.475\textwidth}
\centering
    \subcaptionbox{}
      {\includegraphics[width=0.49\linewidth]{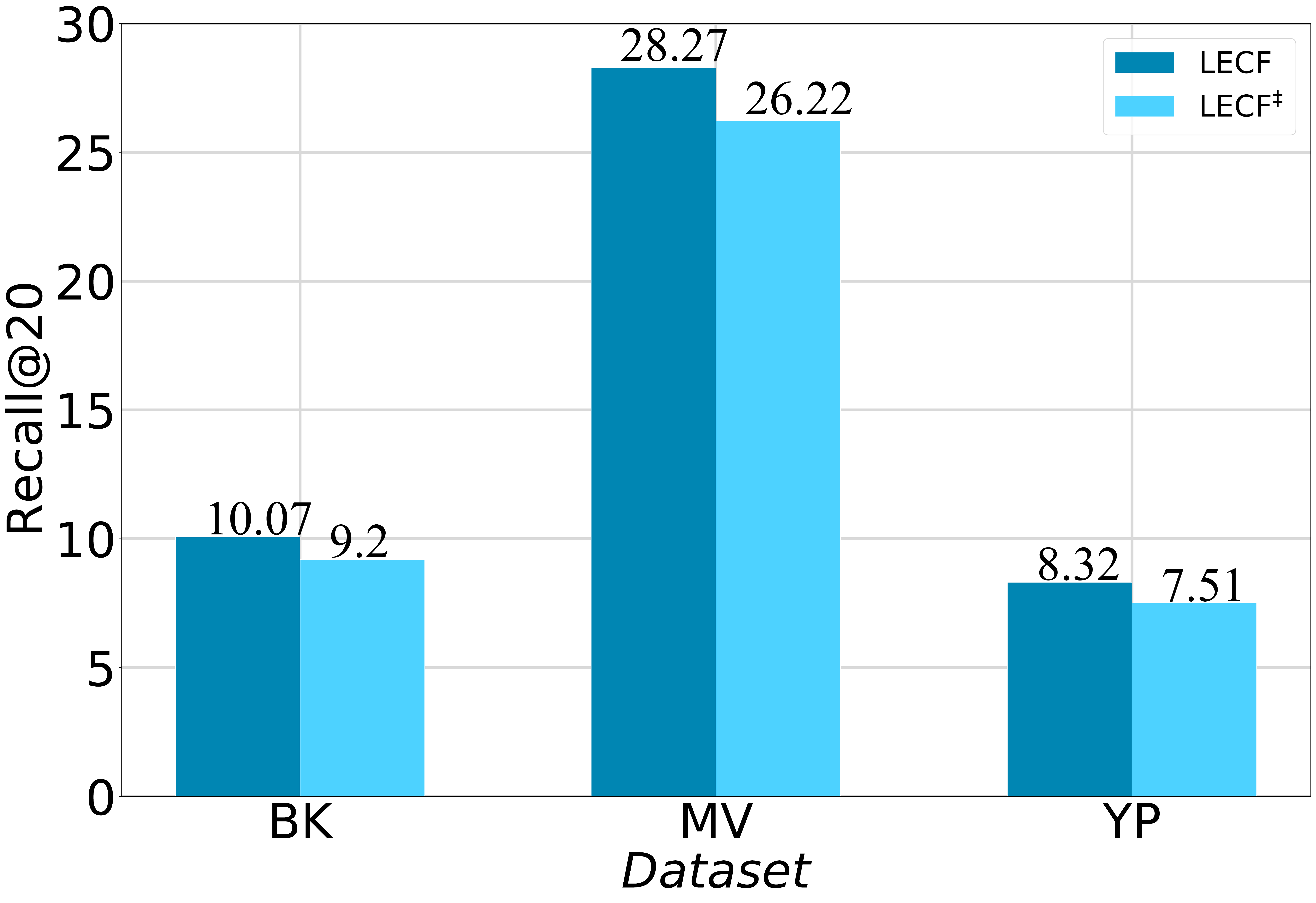}}
    \subcaptionbox{}
      {\includegraphics[width=0.49\linewidth]{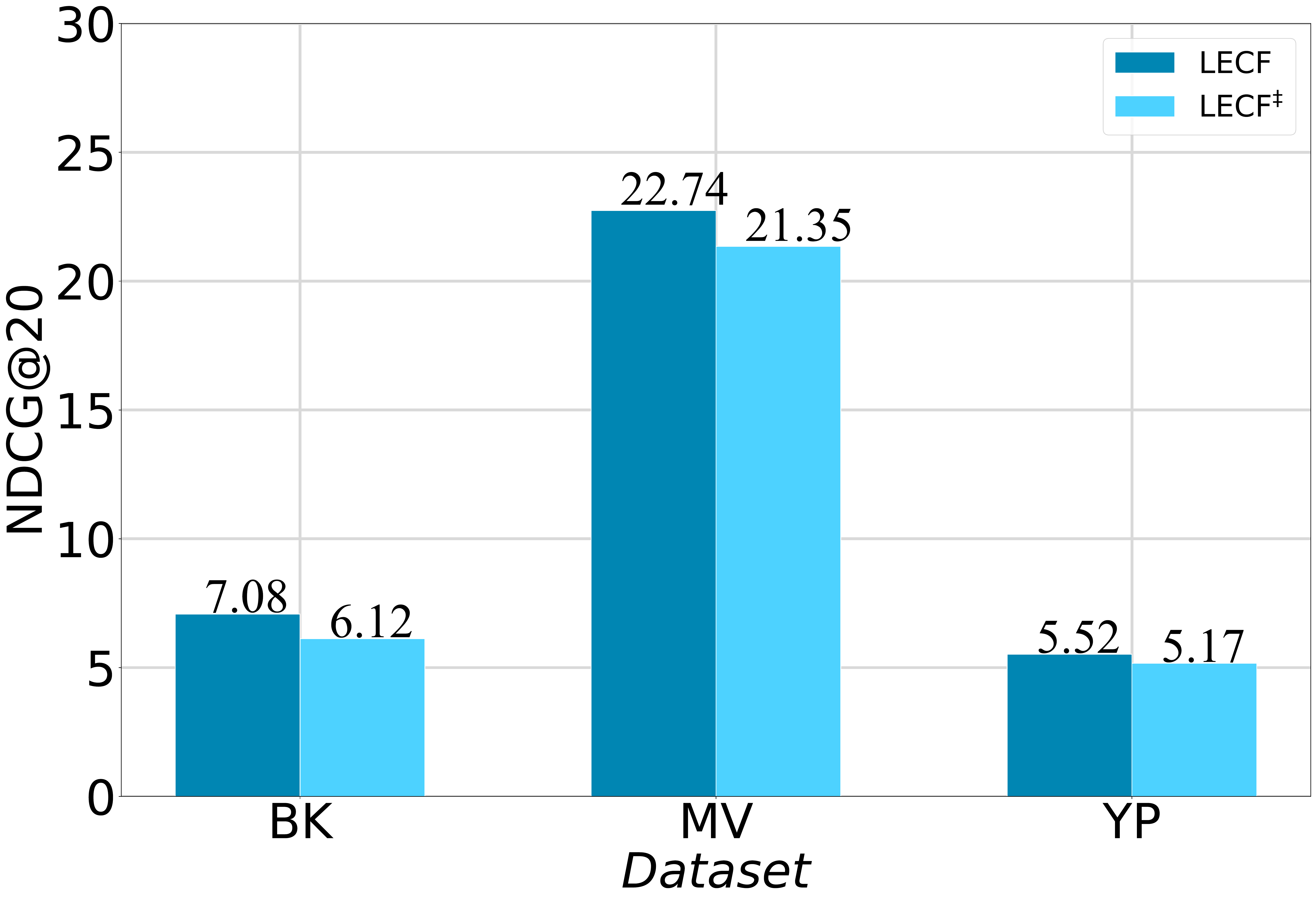}}
  \caption{Ablation experiments of Lorentz equivariance.}\label{fig:ab1}
\end{minipage}
\hspace{0.03\textwidth}
\begin{minipage}[t]{0.475\textwidth}
\centering
    \subcaptionbox{}
      {\includegraphics[width=0.49\linewidth]{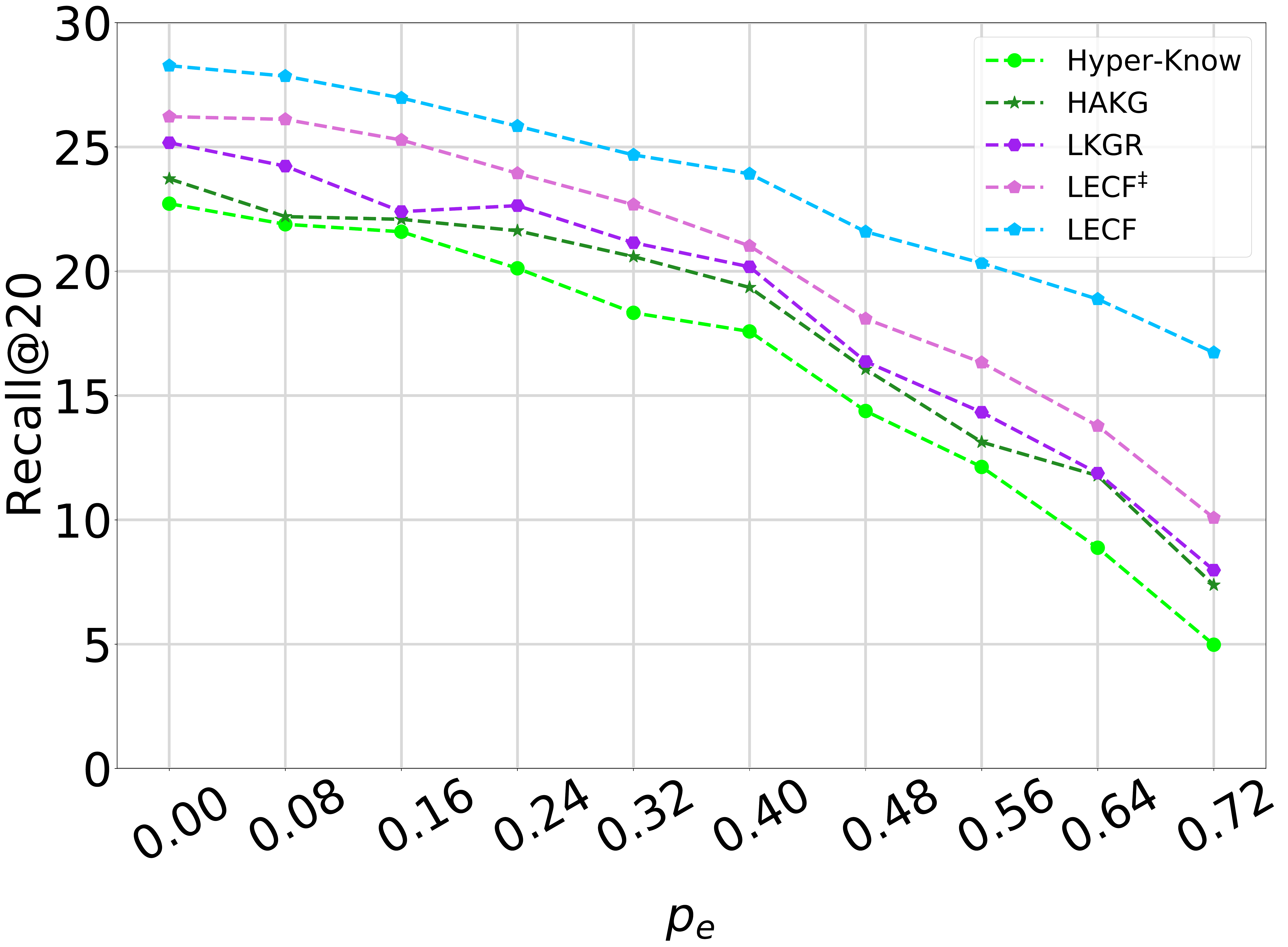}}
    \subcaptionbox{}
      {\includegraphics[width=0.49\linewidth]{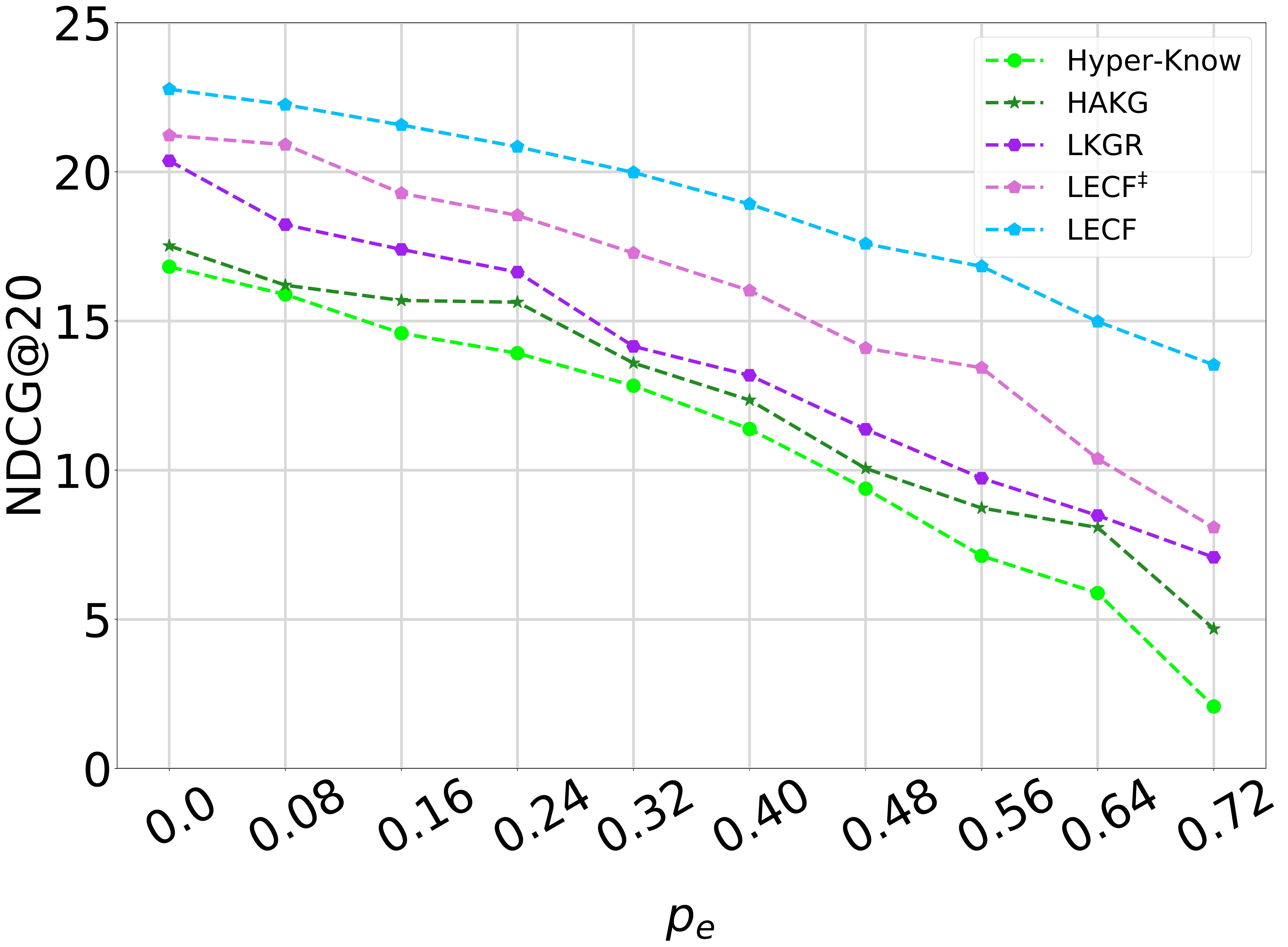}}
  \caption{ Model performance as sparsity increases.}\label{fig:le1}
\end{minipage}
\end{figure}

In this section, we analyze the significance of \lec\ in hyperbolic CF. To demonstrate the effectiveness of preserving \lec, we follow the ablation approach  in \cite{jet2022} to break the \lec\ in LECF by replacing Equation \ref{equ:lecf1} with:
\begin{equation}
    {{\mathbf{m}}_{ui}}={{\phi }_{e}}\left(\mathbf{h}_{u}^{l},\mathbf{h}_{i}^{l},\mathbf{x}_{u}^{l},\mathbf{x}_{i}^{l}\right)
\label{equ:20}
\end{equation}
Since $\mathbf{x}_{u}^{l},\mathbf{x}_{i}^{l}$ are the hyperbolic vectors, transforming them linearly by $\phi_e$ breaks the Lorentz equivariance of the whole model, and this variant is denoted as LECF$^{\ddag}$.  
We performed ablation experiments for  Lorentz equivariance on three datasets. The necessity of enforcing \lec\ in hyperbolic CF is proven since the performance of LECF$^\ddag$ is noticeably worse than that of LECF as shown in Figure \ref{fig:ab1}. We further speculate that the gain produced by the hyperbolic geometry in LECF depends heavily on the fulfillment of \lec.

Moreover, to deeply reveal the effects of \lec, we perform the following two tests (both performed on the MovieLens-20M dataset) to demonstrate the advantages of symmetry-preservation under different situations.

\begin{figure}[ht]
	\centering
		\begin{subfigure}{0.49\linewidth}
			\centering
			\includegraphics[width=\linewidth]{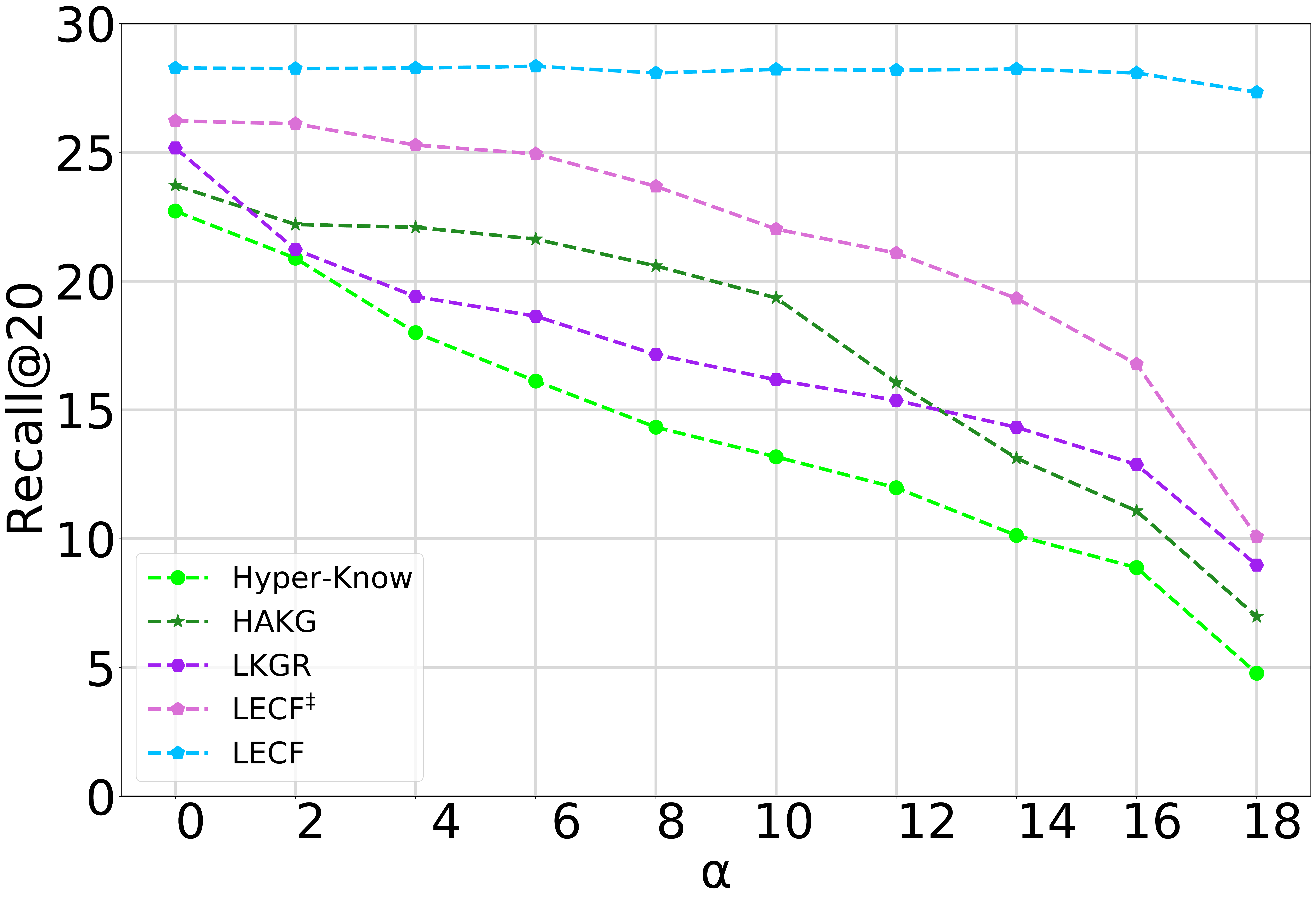}
			\caption{}
			\label{fig:le2a}
		\end{subfigure}
		\begin{subfigure}{0.49\linewidth}
			\centering
			\includegraphics[width=\linewidth]{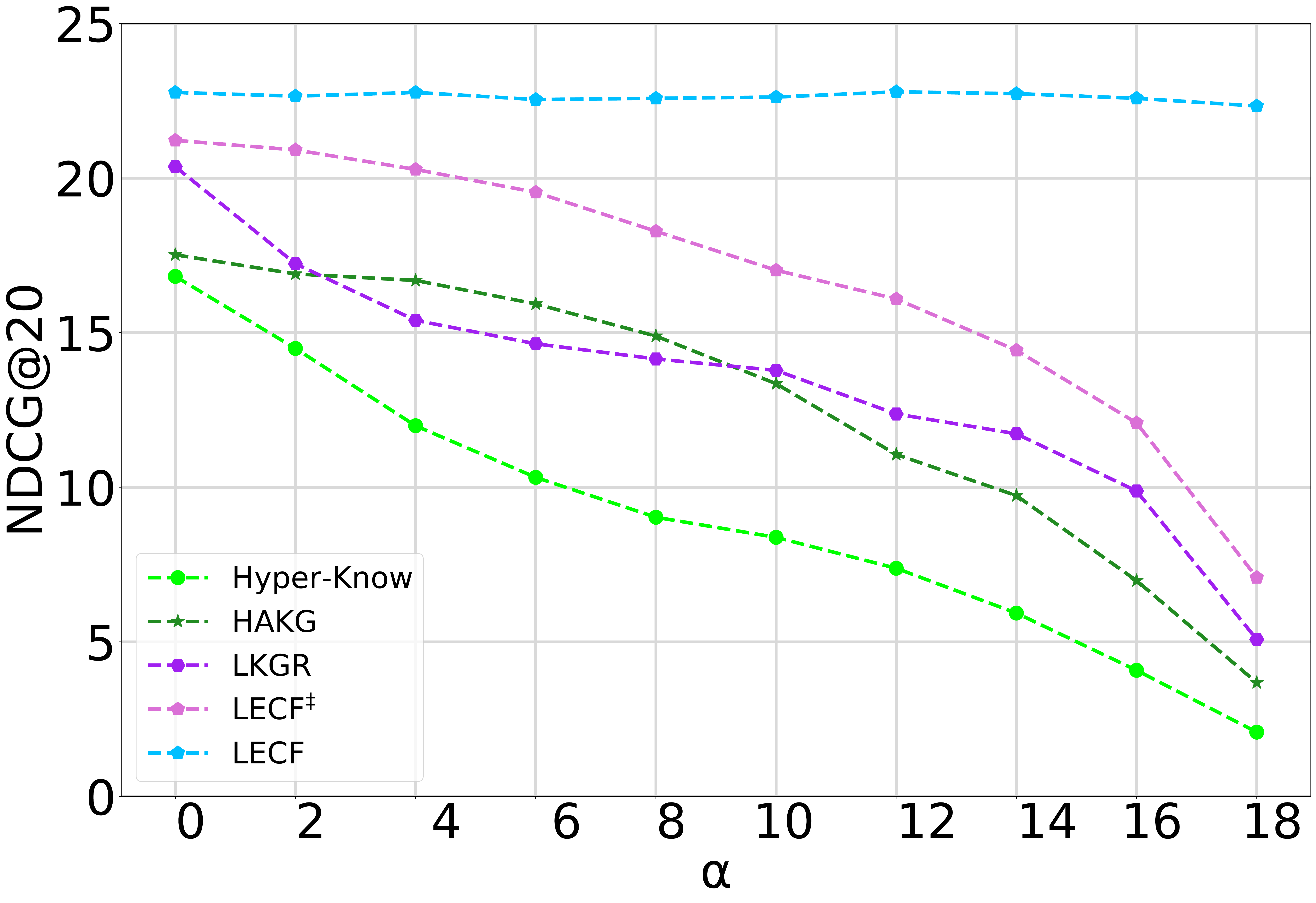}
			\caption{}
			\label{fig:le2b}
		\end{subfigure}
		\begin{subfigure}{0.49\linewidth}
			\centering
			\includegraphics[width=\linewidth]{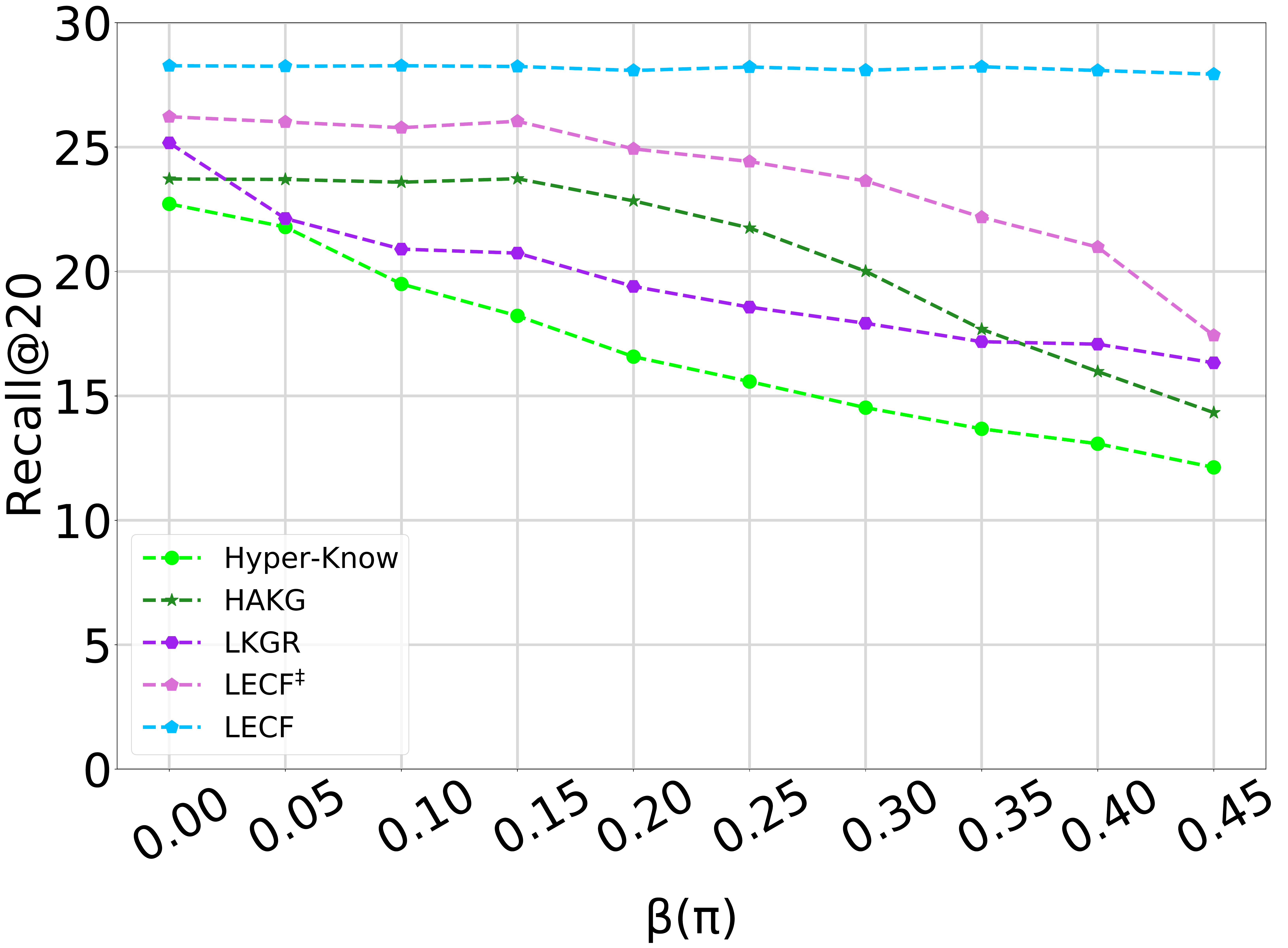}
			\caption{}
			\label{fig:le3a}
		\end{subfigure}
		\begin{subfigure}{0.49\linewidth}
			\centering
			\includegraphics[width=\linewidth]{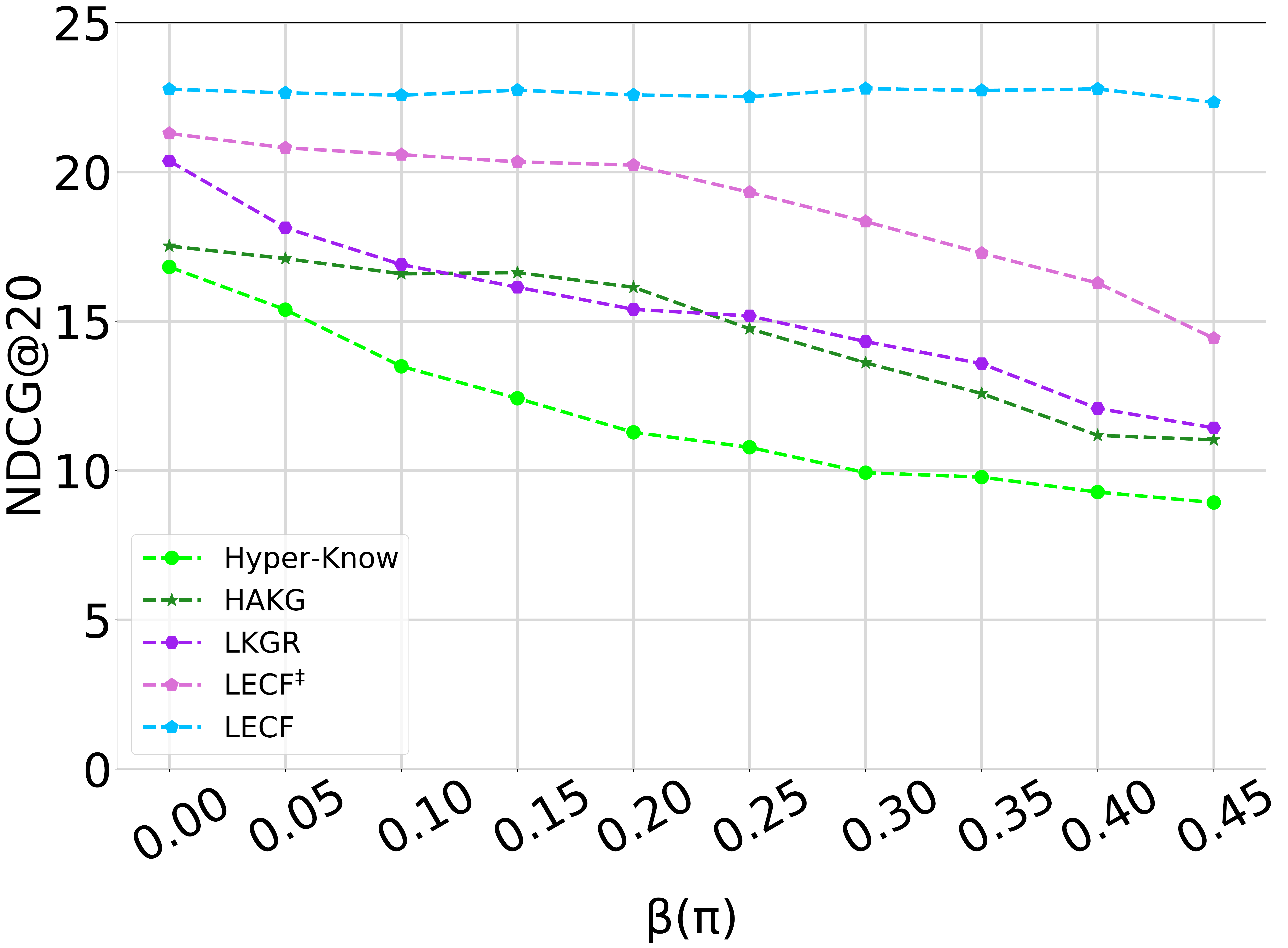}
			\caption{}
			\label{fig:le3b}
		\end{subfigure}
  \caption{Model performance under Lorentz boost (Figure \ref{fig:le2a}, \ref{fig:le2b}) and spatial rotation (Figure \ref{fig:le3a}, \ref{fig:le3b}).}\label{fig:le2}
\end{figure}
\subsubsection{Symmetry Limitation Test.}
In practice, it is difficult to measure the symmetry limitation of a model. However, in very sparse graphs, the impact of symmetry limitation grows dramatically \cite{E(n)GNN2021}.  Therefore, we design the test with a sparse sampling of the original dataset to verify the generalization ability of LECF to symmetric features. Specifically, we eliminate the edges in \uig\ with the ratio of $p_e$, and the experimental results are shown in Figure \ref{fig:le1}. We observe that on both evaluation metrics,  LECF performances drop the slowest as the sparsity of the data increases, while  LECF$^{\ddag}$ does not show particular stability.
\subsubsection{Lorentz Transformation Test.}
Another perspective that supports the significance of symmetry-preserving is the robustness of  \lev\ model under the Lorentz transformations. To better demonstrate this, we perform Lorentz boost and spatial rotation transformation separately on the test data with the training data unchanged. First, the Lorentz boost under the hyperbolic angle $\alpha$ is used to transform the test data, and the results are shown in Figures \ref{fig:le2a}, \ref{fig:le2b}. Second, 
the spatial rotation under angle $\beta$ is performed on the first two dimensions in the space axis, and the results are shown in Figures \ref{fig:le3a}, \ref{fig:le3b}. Overall, Lorentz boost has a greater impact on model performances, which may be related to the geometric characteristics. Accordingly, LECF shows extraordinary stability under both transformations, i.e., the experiments demonstrate that our model perfectly preserves the Lorentz equivariance.

\subsection{ Ablation Study }
We conduct the following ablation studies to obtain deep insights into the effectiveness of all LECF components. 

{\centering
\begin{minipage}[t]{0.45\textwidth}
\captionof{table}{Ablation Study on Top-20 recommendation (\%).}\label{table:ab}
\begin{tabular}{c|ccccc}
\midrule[1.5pt]
 & \multicolumn{1}{c|}{$\otimes$HG}  & \multicolumn{1}{c|}{$\otimes$SA} & \multicolumn{1}{c|}{$\otimes$S1} & \multicolumn{1}{c|}{$\otimes$S2} & \textbf{LECF}  \\ \midrule[0.3pt]
BK-R@20    & 8.50                                             & 9.58                       & 5.28                       & 6.17                       & \textbf{10.07} \\
BK-N@20    & 5.46                                           & 6.75                       & 3.02                       & 5.39                       & \textbf{7.08 } \\\midrule[0.3pt]
MV-R@20    & 20.91                                          & 26.93                      & 16.48                      & 17.04                      & \textbf{28.27} \\
MV-N@20    & 15.99                                         & 22.03                      & 13.60                      & 14.27                      & \textbf{22.74} \\ \midrule[0.3pt]
YP-R@20    & 6.17                                           & 7.39                       & 4.51                       & 5.35                       & \textbf{8.32}  \\
YP-N@20    & 4.78                                          & 5.02                       & 3.97                       & 4.46                       & \textbf{5.52}  \\ \midrule[1.5pt]
\end{tabular}
\begin{tablenotes}
\item[1] \scriptsize{BK, MV, and YP denote
Book-Crossing, MovieLens-20M and Yelp2018 }
\end{tablenotes}\end{minipage}
}

\subsubsection{Effect of hyperbolic geometry.}
To demonstrate the superiority of learning user and item representations in the hyperbolic space over Euclidean space, we transfer all operations in LECF to Euclidean space while retaining all functional components, denoted as LECF$^\mathrm{ \otimes HG}$. In Table \ref{table:ab}, we observe that the results of LECF$^\mathrm{ \otimes HG}$ decline significantly relative to LECF, demonstrating the fundamental significance of hyperbolic geometry for CF based on large-scale networks.

\subsubsection{Effect of mutual signal propagation between the attribute and hyperbolic embeddings.}
In LECF layer, the item attributes, i.e., the entity signals from the KG, are updated mutually with the hyperbolic embeddings of users and items. We cut off these two message propagation paths separately to verify their effectiveness respectively. First, we remove the hyperbolic embeddings in Equation \ref{equ:lecf1} to cut off the signal from \uig\ to KG. We denote this variant as LECF$^\mathrm{ \otimes S1}$. Second, we remove  Lorentz Equivariant Transformation in Equation \ref{equ:lecf2}, whereby the entity signal in 
KG will not be passed to the hyperbolic embeddings of \uig . This variant is denoted as LECF$^\mathrm{\otimes S2}$. The severe deterioration of these two variants demonstrates the importance of bi-directionally propagating information between two graphs.
\subsubsection{Effect of Hyperbolic Sparse Attention Mechanism.}
To perform the corresponding ablation experiment, we first sample the neighbor nodes with a random strategy while setting the attention coefficient to be uniformly distributed. This variant is denoted as LECF$^\mathrm{ \otimes SA}$. For the first time, LECF has successfully implemented a sparse attention mechanism strictly embedded in hyperbolic space, and we can prove its effectiveness from the deterioration of LECF$^\mathrm{ \otimes SA}$.

\begin{figure}[t]
			\centering
\includegraphics[width=\linewidth]{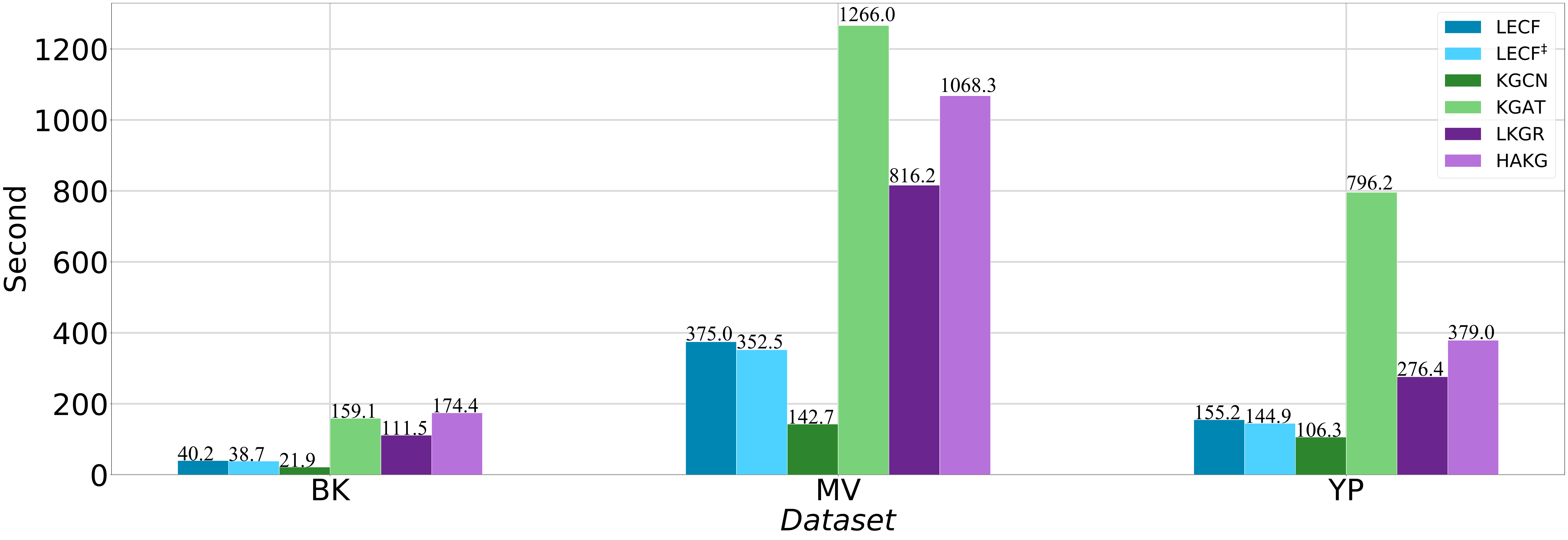}
			\caption{ Training time per epoch (second).}
			\label{fig:trainingtime}
\end{figure}

\subsection{Training Efficiency}
In this section, we compare the training time consumption of each epoch to test the efficiency of the models. All settings for the experiment remain the same as in the previous sections. Figure \ref{fig:trainingtime} shows that the propagation-based method is generally more time-consuming than the regularization-based method. Compared with its hyperbolic peers, LECF is significantly more efficient, which benefits from abandoning the tangent space aggregation and applying the sparse attention mechanism.
\section{Conclusion}

Inspired by numerous explorations of Euclidean equivariance in graph representation learning, we inaugurate the discussion of \lec\ in hyperbolic CF. Since the distance distortion under certain transformations in hyperbolic space is more severe than in Euclidean space, the generalization capability of the hyperbolic CF model depends heavily on the ability to perceive the same characteristics under the Lorentz transformations, that is, to maintain the Lorentz equivariance. In this context, we delicately construct the IAGs and LECF layers that are strictly Lorentz equivariant. 
Specifically, in each IAG,
the proposed Hyperbolic Sparse Attention Mechanisms sample the most informative neighbor nodes to support downstream tasks better. In each LECF layer, the attribute embeddings generated from IAG and the hyperbolic embeddings are mutually updated by the proposed Lorentz Transformation Equivariant in KG so that the high-order entity signals can be passed to the user representation across graphs. Experiments on three public benchmark datasets demonstrate the superiority of LECF. More importantly, we verify that LECF is Lorentz equivariant through extensive experiments and prove that enforcing Lorentz equivariance significantly improves model performance.

\end{document}